\documentclass[12pt]{article}
\usepackage{url}
\input{epsf}

\font\bm=cmmib10 at 10pt
\font\bms=cmmib10 at 7pt \textfont9=\bm \scriptfont9=\bms
\usepackage{amsfonts}
\usepackage{multirow}
\mathchardef\balpha= "790B
\mathchardef\bbeta= "790C
\mathchardef\bTheta= "7902
\mathchardef\bzeta= "7910
\mathchardef\bOmega= "790A
\mathchardef\bGamma= "7900
\mathchardef\bDelta= "7901
\mathchardef\bPhi= "7908
\mathchardef\bphi= "791E
\mathchardef\bomega= "7921
\mathchardef\bxi= "7918
\mathchardef\bet= "7911
\mathchardef\brho= "791A
\mathchardef\btau= "791C
\mathchardef\bmu= "7916
\mathchardef\bvarpi= "7924

\def \sech {\hbox{ sech} }

\def \lvec{(\kern-.26em(}
\def\pmb#1{\setbox0=\hbox{#1}%
\def \lvec{(\kern-.26em(}
\kern-.025em\copy0\kern-\wd0
\kern.05em\copy0\kern-\wd0
\kern-.025em\raise.0433em\box0 }
\mathchardef\btheta= "7912
\usepackage{amsmath}
\usepackage{authblk}

\usepackage{graphicx}
\usepackage{dcolumn}

\begin{document}

\title{Dependence of Earth's Thermal Radiation on Five Most Abundant Greenhouse Gases}
\author[1]{W. A. van Wijngaarden}
\author[2]{W. Happer}
\affil[1]{Department of Physics and Astronomy, York University, Canada, wlaser@yorku.ca}
\affil[2]{Department of Physics, Princeton University, USA, happer@Princeton.edu}
\renewcommand\Affilfont{\itshape\small}
\date{\today}
\maketitle

\noindent The atmospheric temperatures and concentrations of Earth's five most important, greenhouse gases, H$_2$O, CO$_2$, O$_3$, N$_2$O and CH$_4$ control the cloud-free, thermal radiative flux from the Earth to outer space.  Over 1/3 million lines having strengths as low as $10^{-27}$ cm of the HITRAN database were used to evaluate the dependence of the forcing on the gas concentrations.  For a hypothetical, optically thin atmosphere, where there is negligible saturation of the absorption bands, or interference of one type of greenhouse gas with others, the per-molecule forcings are of order $10^{-22}$ W for H$_2$O, CO$_2$, O$_3$, N$_2$O and CH$_4$.  For current atmospheric concentrations, the per-molecule forcings of the abundant greenhouse gases H$_2$O and CO$_2$ are suppressed by four orders of magnitude.  The forcings of the less abundant greenhouse gases, O$_3$, N$_2$O and CH$_4$, are also suppressed, but much less so.  For current concentrations, the per-molecule forcings are two to three orders of magnitude greater for O$_3$, N$_2$O and CH$_4$, than those of H$_2$O or CO$_2$.  Doubling the current concentrations of CO$_2$, N$_2$O or CH$_4$ increases the forcings by a few per cent.  These forcing results are close to previously published values even though the calculations did not utilize either a CO$_2$ or H$_2$O continuum.  The change in surface temperature due to CO$_2$ doubling is estimated taking into account radiative-convective equilibrium of the atmosphere as well as water feedback for the cases of fixed absolute and relative humidities as well as the effect of using a pseudoadiabatic lapse rate to model the troposphere temperature.  Satellite spectral measurements at various latitudes are in excellent quantitative agreement with modelled intensities.  
%
\newpage
\section{Introduction}
The temperature record from 1850 to the present shows the average surface temperature of the Earth has increased by about one degree Celsius\cite{IPCC}.  The Interovernmental Panel on Climate Change (IPCC) attributes most of this temperature rise due to increasing greenhouse gas concentrations associated with anthropogenic activity.  The average concentration of CO$_2$ in the atmosphere has increased from 280 ppm to over 400 ppm largely due to the combustion of fossil fuels.  Concentrations of N$_2$O and CH$_4$ have also risen substantially since the start of the industrial revolution \cite{MaunaLoa}.  

Greenhouse warming of Earth's surface and lower atmosphere is driven by  {\it radiative forcing}, the difference between the flux of thermal radiant energy from a black surface through a hypothetical, transparent atmosphere, and the flux through an atmosphere with greenhouse gases, particulates and clouds, but with the same surface temperature \cite{Schwartz1, Schwartz2}.  This paper examines the effect of greenhouse gas concentrations on thermal radiation for the case of a clear sky.  It considers the five most important naturally occurring greenhouse gases: H$_2$O, CO$_2$, O$_3$, N$_2$O and CH$_4$.  

The spectra of greenhouse gases consists of hundreds of thousands of individual rovibrational spectral lines whose strengths and transition frequencies can be downloaded from the HITRAN database \cite{Rothman92,HITRAN}.  The earliest global warming estimates approximated the multitude of lines by various absorption bands \cite{Chandrasekhar, Manabe1961, Goody}.  The most accurate forcings are found by performing line by line calculations \cite{Edwards1992, Clough1992, Myhre1998, Collins2006, Harde2013, Schreier2014}.  Here, we use line by line calculations to estimate the effects of doubling CO$_2$, N$_2$O and CH$_4$ concentrations from current levels.
The forcings are strongly affected by saturation of the absorption bands and spectral overlap with other greenhouse gases.  Recently, this was found to significantly affect methane forcing \cite{Etminan2016}. 	  

This work downloaded over 1/3 million rovibrational lines from the most recent  HITRAN database to calculate the per-molecule forcings.  The concentration of each greenhouse gas was varied from the optically thin limit where there is negligible saturation or interference of one type of greenhouse gas with others; to current levels where the per-molecule forcings are suppressed by up to four orders of magnitude.  The ``instantaneous" forcings resulting from doubling concentrations of CO$_2$, N$_2$O and CH$_4$ were compared to those published in the literature.  The change to the temperature profile was estimated taking into account radiative-conective equilibrium of the atmosphere as well as water feedback.  Finally, the intensities modelled at the top of the atmosphere were compared to satellite spectral measurements at various latitudes.   

\begin{figure}[t]
\includegraphics[height=100mm,width=1\columnwidth]{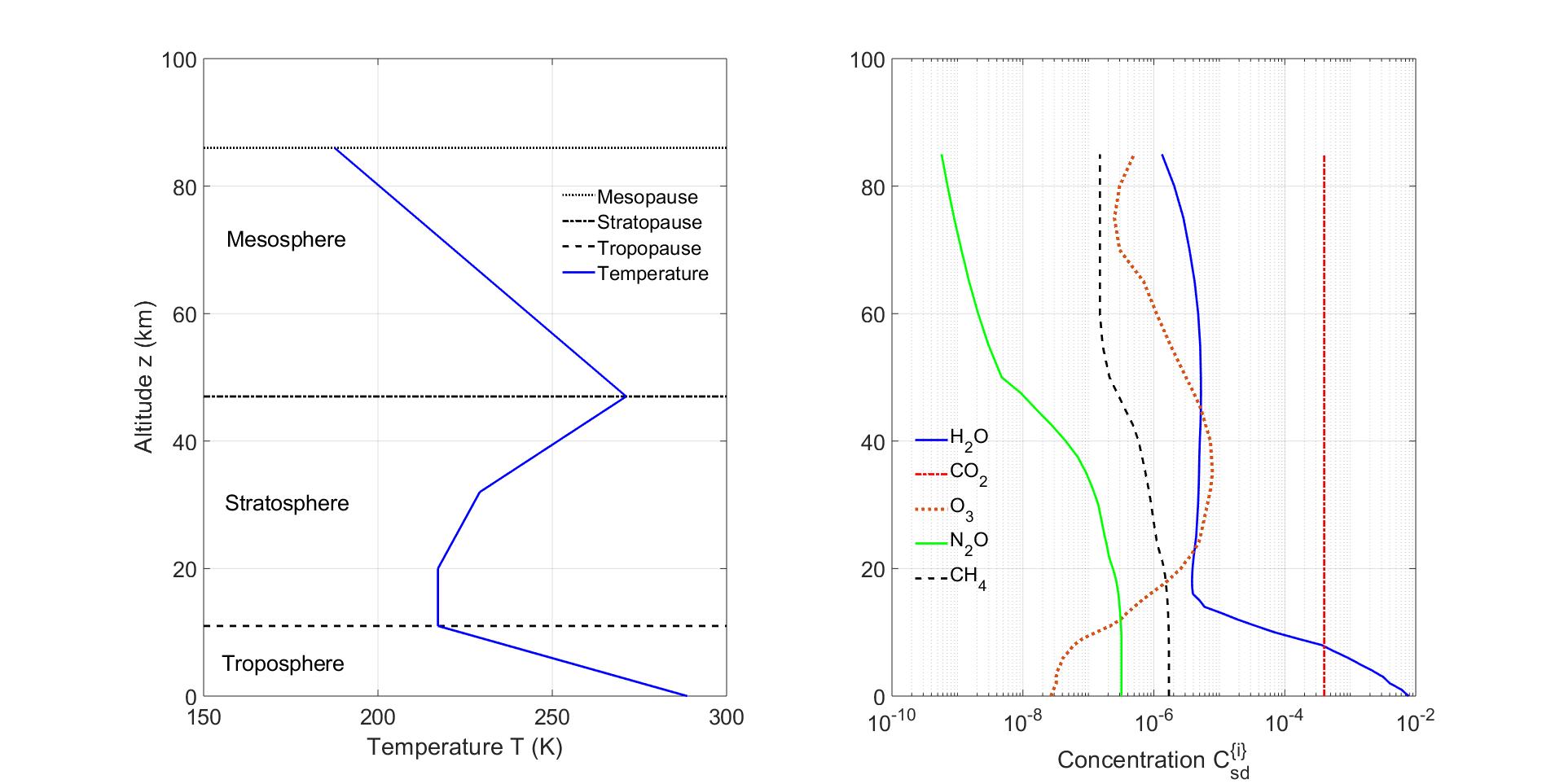}
\caption{{\bf Left.} A standard atmospheric temperature profile\cite{Temp}, $T=T(z)$. The Earth's mean surface temperature i $T(0) = 288.7$ K . {\bf Right.} Standard concentrations\cite{Anderson}, $C^{\{i\}}_{\rm sd}$ for greenhouse molecules versus altitude $z$. 
\label{GGNT}}
\end{figure}

\section{Altitude Profiles of Temperature and Greenhouse Gases \label{he}}

Radiation transfer in the cloud-free atmosphere of the Earth is controlled by the temperature $T=T(z)$ at the altitude $z$ and the number densities, $N^{\{i\}}=N^{\{i\}}(z)$ of the $i$th type of molecule.  
Representative midlatitude altitude profiles of temperature \cite{Temp}, and greenhouse gas concentrations \cite{Anderson}, are shown in Fig. \ref{GGNT}.  Altitude profiles directly measured by radiosondes in ascending balloons are always more complicated than those of Fig. \ref{GGNT}, which can be thought of as appropriate average profiles for the year 2020.  

We divided the atmosphere into 500 altitude segments, 100 segments for each layer:  the troposphere, lower stratosphere, mid stratosphere, upper stratosphere and mesosphere.  The segment midpoints are labeled by the integers $i = 1, 2, 3, ..., 500$.  We characterize the initial temperature profile of the atmosphere, with six breakpoints, with temperatures $\theta_{\alpha}$ and altitudes $\zeta_{\alpha}$, where $\alpha = 0, 1, ..., 5$.  Between the breakpoints, the atmosphere is assumed to have constant temperature lapse rates $L_{\alpha}= -dT/dz = -(\theta_{\alpha}-\theta_{\alpha-1})/(\zeta_{\alpha}-\zeta_{\alpha-1})$.  For a midlatitude standard atmosphere, the breakpoint temperatures and altitudes are
	
\begin{equation}
{\bf \theta}=\left[\begin{array}{r}  288.7\\ 217.2\\ 217.2\\ 229.2\\ 271.2\\ 187.5\end{array}\right]\hbox{ K},\quad\hbox{and}\quad
{\bf \zeta}=\left[\begin{array}{r} 0\\ 11\\20 \\32\\ 47\\86 \end{array}\right]\hbox{km.}
\label{es1}
\end{equation}
	
\noindent The lapse rates are $L =[\,6.5,0,-1,-2.8,2.1\,]$ K km$^{-1}$.  The temperature profile determined by (1) is shown as the solid blue line in the left panel of Fig. 1.

In the troposphere, convective transport of sensible and latent heat of water vapor, especially near the equator, is as important as radiant heat transfer.  Above the troposphere is the stratosphere, which extends from the tropopause to the stratopause, at a typical altitude of $z_{\rm sp} = 47$ km, as shown in Fig. \ref{GGNT}.  The temperature in the stratosphere is nearly constant at low altitudes, but increases at higher altitudes due to the heating of ozone molecules that absorb ultraviolet sunlight.  The stratosphere is much more stable to vertical displacements of air parcels than the troposphere and negligible moist convection occurs because of the very low water vapor concentration.

Above the stratosphere is the mesosphere, which extends from the stratopause to the mesopause at an altitude of about  $z_{\rm mp} = 86$ km.  With increasing altitudes, radiative cooling, mainly by CO$_2$, becomes increasingly more important compared to heating by solar ultraviolet radiation.  This causes the temperature to decrease with increasing altitude in the mesosphere.

The vertical radiation flux changes rapidly in the troposphere and stratosphere compared to the mesosphere where the atmospheric density is very low.  Changes in flux above the mesopause are negligible and the mesopause is therefore referred to as ``the top of the atmosphere" (TOA), with respect to radiation transfer.

\begin{table}
\begin{center}
\begin{tabular}{|c|c| c|}
 \hline
 $i$&Molecule& $\hat N^{\{i\}}_{\rm sd}$ (cm$^{-2}$)\\ [0.5ex]
 \hline\hline
 1& H$_2$O &$4.67\times 10^{22}$\\
 \hline
 2& CO$_2$ &$8.61\times 10^{21}$\\
 \hline
 3&O$_3$   &$9.22\times 10^{18}$\\
 \hline
 4&N$_2$O  &$6.61\times 10^{18}$\\
 \hline
 5&CH$_4$  &$3.76\times 10^{19}$\\
 \hline
 \end{tabular}
\end{center}
\caption{Column densities, $\hat N^{\{i\}}_{\rm sd}$, of the 5 most abundant greenhouse gases obtained using the standard altitudinal profiles \cite{Anderson} of Fig. \ref{GGNT}.
\label{acd}}
\end{table}

The standard concentrations for the ith greenhouse gas, $C^{\{i\}}_{\rm sd}$, based on observations\cite{Anderson}, are shown as functions of altitude on the right of Fig. \ref{GGNT}.  The sea level concentrations are $7,750$ ppm of H$_2$O, $1.8$ ppm of CH$_4$ and $0.32$ ppm of N$_2$O. The O$_3$ concentration peaks at $7.8$ ppm at an altitude of 35 km, and the CO$_2$ concentration was $400$ ppm at all altitudes.  Integrating the concentrations over an atmospheric column having a cross sectional area of 1 cm$^2$ yields the column number density of the $i$th type of molecule $\hat N^{\{i\}}_{\rm sd}$ which are listed in Table \ref{acd}.

\section {Greenhouse Gas Lines}

A line by line calculation of radiative forcing utilizes various  parameters that are now briefly discussed.  

\subsection{Line Intensities}

\begin{figure}[t]
\includegraphics[height=100mm,width=.95\columnwidth]{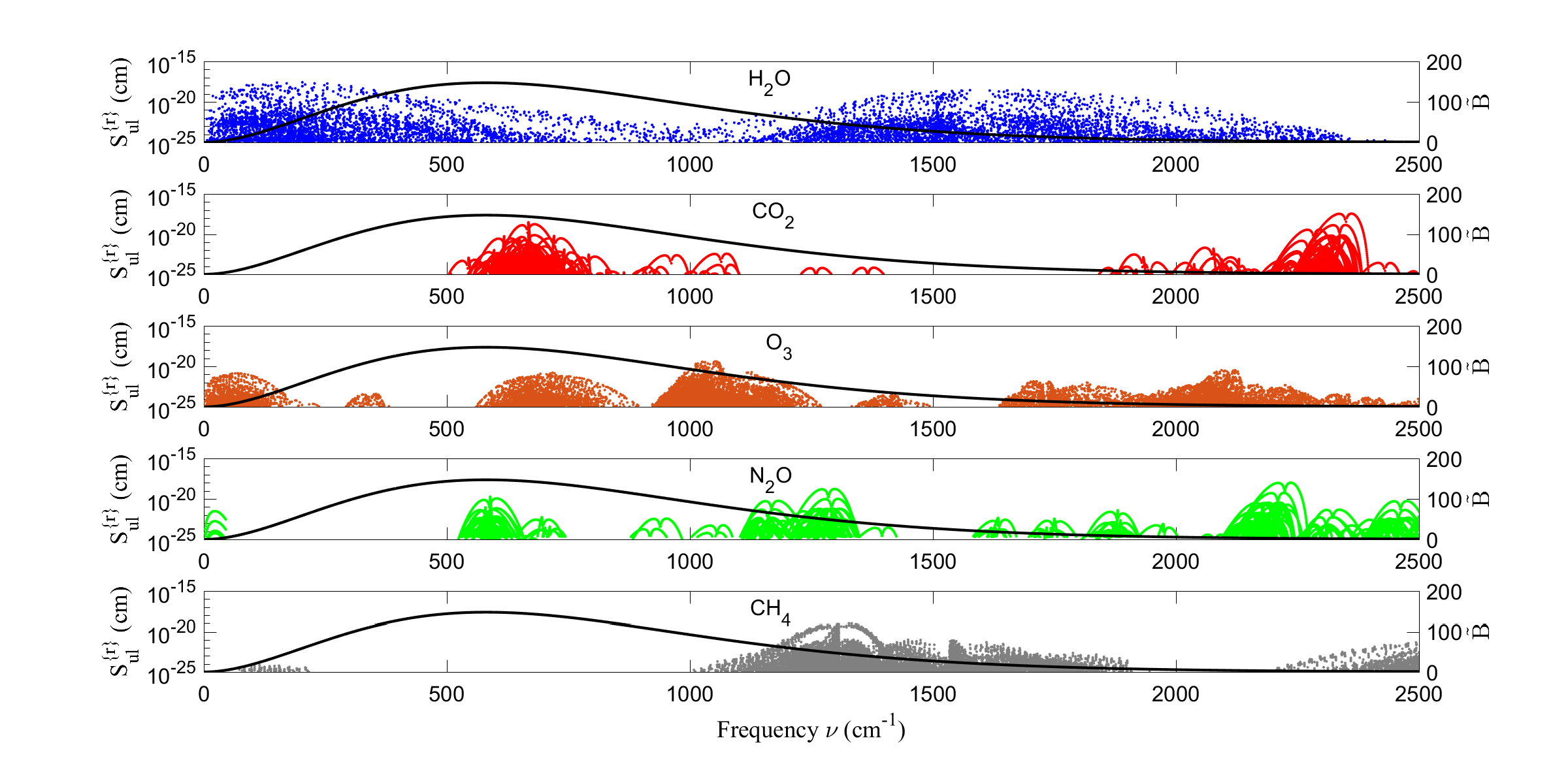}
\caption{Reference line intensities, $S_{ul}^{\{r\}}$ of (\ref{lbl36}) for H$_2$O, CO$_2$, O$_3$, N$_2$O and  CH$_4$ from the HITRAN data base \cite{HITRAN}.  The  horizontal coordinate of each point represents the Bohr frequency $\nu_{ul}$ of a transition from an upper level $u$ to a lower level $l$.  The vertical coordinate of the point is the line intensity.  For greater clarity we have plotted only 1/10, chosen at random, of the extremely large number of O$_3$ line intensities.  The numbers of lines (in parenthesis) used for this work were: H$_2$O (31,112), CO$_2$ (20,569), O$_3$ (210,295), N$_2$O (43,152) and CH$_4$ (43,696).  The smooth line is the Planck spectral intensity, $\tilde B$ of (\ref{b12}) in units of mW cm m$^{-2}$ sr$^{-1}$ for the HITRAN reference temperature, $T^{\{r\}} = 296$ K.}
\label{LI}
\end{figure}

Fig. \ref{LI} illustrates the greenhouse gas lines considered in this work.  The  Bohr frequency $\nu_{ul}$ for a radiative transition from a lower level $l$ of energy $E_l$ to an upper level $u$ of energy $E_u$ of the same molecule is denoted by

\begin{equation}
\nu_{ul}=\frac{E_{ul}}{h c},\quad\hbox{where}\quad E_{ul}= E_u-E_l.
\label{lbl2}
\end{equation}

\noindent where the energy of a resonant photon is $E_{ul}$, $h$ is Planck's constant and $c$ is the speed of light.

The cross section, $\sigma^{\{i\}}=\sigma$, for the $i$th type of greenhouse molecule is written as the sum of partial cross sections $\sigma_{ul}$, corresponding to each Bohr frequency $\nu_{ul}$,

\begin{equation}
\sigma=\sum_{ul} \sigma_{ul}.
\label{lbl16}
\end{equation}

\noindent The partial cross section, $\sigma_{ul}$, is assumed to be the product of a  lineshape function, $G_{ul}=G_{ul}(\nu,\tau)$, and a
line intensity, $S_{ul}=S_{ul}(T)$,
\begin{equation}
\sigma_{ul}=G_{ul}S_{ul}.
\label{lbl18}
\end{equation}

\noindent The lineshape functions, $G_{ul}$, are normalized to have unit area,

\begin{equation}
\int_0^{\infty}G_{ul}d\nu=1
\label{lbl20}
\end{equation}

\noindent and have units of cm.  The line intensity is

\begin{equation}
S_{ul} =\eta_{u}\pi r_e f_{ul} W_l\left(1-e^{-\nu_{ul}/\nu_T}\right)=\frac{\eta_u W_u\Gamma_{ul}E_{ul}}{4\pi \tilde B_{ul}}.
\label{lbl24}
\end{equation}

\noindent In (\ref{lbl24}), $S_{ul}$ has the units of cm, $r_e=e^2/(m_e c^2)$ is the classical electron radius, where $e$ is the elementary charge and $m_e$ is the electron mass.  The isotopologue fractions are $\eta_u$.  For the most abundant isotopologues of CO$_2$,

\begin{equation}
\eta_{u}=\left \{\begin{array}{rr} 0.9843&\hbox{ for $^{16}$O $^{12}$C $^{16}$O}\\
0.0110 &\hbox{ for $^{16}$O $^{13}$C $^{16}$O}\\
0.0040 &\hbox{ for $^{16}$O $^{12}$C $^{18}$O}\\
0.0007 &\hbox{ for $^{16}$O $^{12}$C $^{17}$O}. \end{array}\right.
\label{lbl4}
\end{equation}

\noindent The last term of (\ref{lbl24}) contains the spectral Planck intensity evaluated at the frequency $\nu_{ul}$,

\begin{equation}
\tilde B_{ul}=\tilde B(\nu_{ul},T).
\label{lbl32}
\end{equation}

\noindent The Planck intensity is given by 
 
\begin{equation}
\tilde B(\nu,T)=\frac{2h c^2\nu^3}{e^{\nu c\, h/(k_{\rm B}T)}-1}
\label{b12}
\end{equation}

\noindent The radiation frequency, $\nu=1/\lambda$ is the inverse of the wavelength $\lambda$ and has units of cm$^{-1}$.  

\noindent The probability $W_n$ (with $n=u$ or $n=l$) to find a molecule in the rovibrational level $n$ is

\begin{equation}
W_n =\frac{g_n e^{-E_n/k_{\rm B}T}}{Q}.
\label{lbl26}
\end{equation}

\noindent Here $g_n$ is the statistical weight of the level $n$, the number of independent quantum states with the same energy $E_n$.
For molecules in the level $n$, the statistical weight can be taken to be
\begin{equation}
g_n=(2j_n+1)k_n,
\label{lbl27}
\end{equation}
where $j_n$ is the rotational angular momentum quantum number, and $k_n$ is the nuclear degeneracy factor, that depends on whether the spins of the nuclei are identical or not.  The partition function, $Q=Q(T)$, of the molecule is
\begin{equation}
Q=\sum_n g_n e^{-E_n/k_{\rm B}T}.
\label{lbl28}
\end{equation}
The oscillator strength, $f_{ul}$, of (\ref{lbl24}) is related to the matrix elements of the electric dipole moment {\bf M} of the molecule,  between the upper energy basis state $|j_u m_u\rangle$ with azimuthal quantum number $m_u$ and the lower energy basis state $|j_l m_l\rangle$, by
\begin{equation}
f_{ul}=\frac{4\pi\nu_{ul}}{3 g_lc\,r_e \hbar}
\sum_{m_u m_l}\langle u \,m_u|{\bf M}|l\, m_l\rangle\cdot\langle l\, m_l|{\bf M}|u\, m_u\rangle.
\label{lbl30}
\end{equation}
The quantum numbers $m_u$ label the various degenerate substates of the upper level $u$ and the $m_l$ label the substates of the lower level $l$.  If the levels are characterized by rotational quantum numbers $j_u$ and $j_l$, the quantum numbers $m_u$ and $m_l$ can be thought of as the corresponding azimuthal quantum numbers, for example, $m_u=j_u, j_u-1,\ldots,-j_u$.

The rate of spontaneous emission of photons when the molecule makes transitions from the upper level $u$ to the lower level $l$ is $\Gamma_{ul}$, the same as the Einstein $A$ coefficient, $\Gamma_{ul}= A_{ul}$ \cite{Einstein, Dirac1930}. The spontaneous emission rate is related to the oscillator strength by
\begin{equation}
\Gamma_{ul}=\frac{8\pi^2c\,r_e \nu_{ul}^2f_{ul}g_l}{ g_u}.
\label{lbl31}
\end{equation}

From inspection of (\ref{lbl24}) we see that the line intensity $S_{ul}=S_{ul}(T)$ at some arbitrary temperature $T$ is related to the intensity, $S_{ul}^{\{r\}}=S_{ul}(T^{\{r\}})$ at a reference temperature $T^{\{r\}}$ where the partition function of (\ref{lbl28}) is related to $Q^{\{r\}}=Q(T^{\{r\}})$  by
\begin{equation}
S_{ul} =S_{ul}^{\{r\}}\frac{Q^{\{r\}}}{Q}\left(\frac{e^{-E_l/k_BT}}{e^{-E_l/k_{\rm B} T^{\{r\}}}}\right)\left(\frac{1-e^{-\nu_{ul}/\nu_T}}{1-e^{-\nu_{ul}/\nu_{T^{\{r\}}}}}\right).
\label{lbl36}
\end{equation}

\noindent This work considered all lines in the HITRAN database of the five gases under consideration in having intensities greater than $10^{-25}$ cm.  For H$_2$O, lines having intensities greater than $10^{-27}$ cm were included since water vapor has an order of magnitude greater density than any other greenhouse gas near the Earth's surface. 
\subsection{Lineshapes}
It is convenient to write the lineshape function as the product of a ``core" profile  $C_{ul}=C_{ul}(\nu)$ and a wing-suppression factor, $\chi_{ul}=\chi_{ul}(\nu)$,
\begin{equation}
G_{ul}=C_{ul}\chi_{ul}.
\label{lsh4}
\end{equation}
\paragraph{Core functions} The core function is normally taken to be a Doppler broadened Lorentzian, often called a Voigt profile\cite{Voigt12}. If we average over a Maxwellian distribution of velocities $v$ for molecules of mass $m$ at the temperature $T$ we find that the core function has the form
\begin{eqnarray}
C_{ul}&=&\frac{\mu_{ul}}{\pi}\sqrt{\frac{m}{2\pi k_{\rm B} T}}\int_{-\infty}^{\infty}\frac{e^{-mv^2/2
		k_{\rm B}T}dv}{\mu_{ul}^2+(\nu-\nu_{ul}[1+v/c])^2}\nonumber\\
\label{lsh6}
\end{eqnarray}
The half width at half maximum of a purely Doppler broadened line is
\begin{equation}
\Delta \nu_{ul}=\nu_{ul}\sqrt{\frac{2 k_{\rm B} T\ln 2}{mc^2}}\approx 0.0005 \hbox{ cm}^{-1}
\label{lsh7}
\end{equation}
Here, the representative Doppler half width is for a resonance frequency $\nu_{ul}$ of the 667 cm$^{-1}$ band of a CO$_2$ molecule near the cold mesopause.

The half width at half maximum, $\mu_{ul}$, of the Lorentzian function in (\ref{lsh6}) is almost entirely due to collisions.  The contribution to $\mu_{ul}$ from spontaneous radiative decay is negligible at altitudes below the mesopause.  For the bending mode of CO$_2$, representative values\cite{Rothman92} of $\mu_{ul}$ at atmospheric pressure $p$ are
\begin{equation}
\mu_{ul} \approx 0.07 \frac{p}{p_0} \hbox{ cm}^{-1}.
\label{lsh8}
\end{equation}
Here $p_0= 1$ bar, the approximate atmospheric pressure at mean sea level.  The pressure broadening coefficients depend somewhat on temperature and on the particular resonance, $ul$, involved.

For (\ref{lsh6}), a small, temperature dependent pressure shift\cite{Rothman92} must be added to the free molecule Bohr frequency  of (\ref{lbl2}), which we denote by  $\nu_{ul}^{\{0\}}$, to define the resonance frequency
\begin{equation}
\nu_{ul}= \nu_{ul}^{(0)}+\delta_{ul}\, p/p_0.
\label{lsh9a}
\end{equation}
The magnitude of the pressure shift coefficient $\delta_{ul}$ is of order
\begin{equation}
|\delta_{ul}|\approx 0.001\hbox{ cm}^{-1},
\label{lsh9b}
\end{equation}
comparable to the Doppler half width (\ref{lsh7}).
The small pressure shifts have negligible influence on radiative forcing calculations, but the pressure broadening coefficients of (\ref{lsh8}) have a large effect.  Collisions significantly broaden absorption lines in the troposphere and stratosphere.

\paragraph{Wing-suppression functions}
Lorentz profiles give far too much absorption for large detunings $|\nu-\nu_{ul}|$, so it is necessary to include wing suppression factors $\chi$ in the expression (\ref{lsh4}) for the lineshape function \cite{Schwarzkopf85, Lacis91}.  Lorentzian lineshapes result from assuming an infinitely short collision duration, but in fact collisions take a few ps for completion.  The collisional interactions that lead to wing suppression are not known well enough for reliable theoretical calculations, so we use the empirical wing suppression factor
\begin{equation}	
\chi_{ul}(\nu)=\sech^2([\nu-\nu_{ul}]/\varpi).
\label{pf2}
\end{equation}
Measurements on bands of CO$_2$, for example by Edwards and Strow \cite{Edwards1991}, suggest that the far wings decrease approximately exponentially with detuning, $|\nu-\nu_{ul}|$, as does the wing suppression function (\ref{pf2}).  We used the width parameter $\varpi= 2$ cm$^{-1}$, corresponding to a collision duration of a few picoseconds.

\section{Radiation}

In cloud-free air where scattering is negligible, radiation transport is governed by the Schwarzschild equation \cite{Schwarzschild1906},

\begin{equation}
\cos \theta \frac{\partial \tilde I}{\partial \tau}=-(\tilde I-\tilde B)
\label{nsa4}
\end{equation}

\noindent where ${\tilde I} = {\tilde I}(\nu,z,\theta)$ is the spectral intensity of a pencil of radiation of frequency between $\nu$ and $\nu + d\nu$ at altitude $z$.  The pencil makes an angle $\theta$ to the vertical.  In thermal equilibrium, the spectral intensity $\tilde I$ equals the Planck intensity given by (\ref{b12}).  The optical depth is defined by 

\begin{equation}
\tau(z,\nu)=\int_0^{z}dz' \kappa(z',\nu),
\label{b10}
\end{equation}

\noindent where the net attenuation coefficient due to molecules absorbing and remitting light of frequency $\nu$ at altitude $z$ is given by 

\begin{equation}
\kappa(z,\nu)=\sum_i N^{\{i\}}(z)\sigma^{\{i\}}(z,\nu).
\label{b8}
\end{equation}

\noindent Here $N^{\{i\}}(z)$ is the density of greenhouse gas molecule of type $i$ and $\sigma^{\{i\}}=\sigma^{\{i\}}(z,\nu)$ is its absorption cross section for radiation of frequency $\nu$ at the altitude $z$ given by (\ref{lbl16}).  The cross section can depend strongly on altitude because temperature and pressure are functions of altitude.  Temperature controls the distribution of the molecules between translational, rotational and vibrational states.  Pressure, together with temperature, determines the width of the molecular resonance lines.  

The optical depth from the surface to the top of the radiative atmosphere, the altitude $z_{\rm mp}$ of the mesopause, is

\begin{equation}
\tau_{\infty}=\tau_{\rm mp}=\int_0^{z_{\rm mp}}dz'\kappa(z',\nu).
\label{b11}
\end{equation}

\noindent As indicated by the notation (\ref{b11}), we have assumed that the optical depth $\tau_{\rm mp}$ at the mesopause altitude $z_{\rm mp}$ differs negligibly from the optical depth $\tau_{\infty}$ at infinite altitude since there is so little opacity of the atmosphere above the mesopause.

The Schwarzschild equation (\ref{nsa4}) can be solved to find the intensity \cite{Buglia,Rodgers1966}

\begin{eqnarray}
\hbox{For $\varsigma>0$}:\quad\tilde I(\tau,\varsigma)&=& +\varsigma \int_0^{\tau}d\tau' e^{-\varsigma (\tau-\tau')}\tilde B(\tau')+e^{-\varsigma\tau}\tilde I(0,\varsigma)\label{vn48}\\
\hbox{For $\varsigma<0$}:\quad\tilde I(\tau,\varsigma)&=& -\varsigma \int_{\tau}^{\tau_{\infty}}d\tau' e^{-\varsigma (\tau-\tau')}\tilde B(\tau')\label{vn50}
\end{eqnarray}

\noindent where $\varsigma= \sec \theta$.  For simplicity, we assume the surface intensity is the product of $B_s=B(T_s)$, the Planck intensity (\ref{b12}) for a temperature $T_s$, and an angle independent emissivity $\epsilon_s=\epsilon_s(\nu)$,

\begin{equation}
\tilde I(0,\varsigma)= \epsilon_s\tilde B_s.
\label{vna51}
\end{equation}

\noindent Over most of the Earth's surface the thermal infrared emissivity $\epsilon_s$, is observed to be in the interval  $[0.9\,<\epsilon_s\,<1]$ \cite{Wilber}. Negligible error is introduced by setting $\epsilon_s=1$ in spectral regions of high atmospheric opacity, with $\tau_{\infty}>>1$.
With only radiative heat transfer from the surface and with negligible conduction or convection, there can be a discontinuous jump between the surface temperature $T_s$ and the temperature $T_0$ of the air just above.  For the rest of this paper we assume that any temperature jump is negligibly small, and $T_s=T_0$ and $\tilde B_s=\tilde B_0$.

For discussions of greenhouse gas forcing, we will be particularly interested in the upward flux

\begin{equation}
\tilde Z=\int_{4\pi}d\Omega\,\cos\theta\,\tilde I.
\label{b5}
\end{equation}

\noindent Using (27) and (28) in (30), we find

\begin{eqnarray}
{{\tilde Z}\over {2 \pi}}&=&\int_0^{\tau}d\tau'E_2(\tau-\tau')\tilde B(\tau') - \int_{\tau}^{\tau_{\infty}} d\tau' E_2(\tau'-\tau)\tilde B(\tau')+ \epsilon_s\tilde B_sE_3(\tau)  \nonumber\\
&=&-\int_0^{\tau_{\infty}} d\tau' E_3(\vert \tau - \tau' \vert) {{\partial \tilde B(\tau')}\over{\partial \tau'}} +
\tilde B(\tau_{\infty}) E_3(\tau_{\infty} - \tau) +[\epsilon_s \tilde B_s - \tilde B(0)] E_3(\tau).
\label{vn14}
\end{eqnarray}

\noindent In this work, we have assumed that $\epsilon_s\tilde B_s = \tilde B(0)$.
Equation (31) is the fundamental expression for the net upward flux in an atmosphere with negligible scattering and has been known for a long time. For example, the first line can be found in the NASA reports:  Equation (2a) of Yoshikawa \cite{Yoshikawa}, or Equation (3.19) of Buglia \cite{Buglia}.  Equation (\ref{vn14}) contains exponential-integral functions, $E_n(\tau)$, that account for slant paths of radiation between different altitudes.  They are defined for integers $n=1,2,3,\ldots$ by
\begin{equation}
E_n(\tau)=\int_1^{\infty}d\varsigma\,\varsigma^{-n}\,e^{-\varsigma\tau},
\label{b40}
\end{equation}
as discussed in Appendix 1 of Chandrasekhar \cite{Chandrasekhar}, or Section {5.1.4} of Abramowitz and Stegun \cite{Abramowitz}.

For frequencies where the atmosphere is not too optically thick, with $\tau_{\infty} \le 10$, we evaluate $\tilde Z$ from the first line of (\ref{vn14}).  Fast Fourier transforms are used to calculate a discretized version of the convolution with $\tilde B$.  For frequencies where the atmosphere is thicker, we use the expression on the first line of (\ref{vn14}) to calculate $\tilde Z$ for 10 e-foldings of optical depth down from the top of the atmosphere, $\tau_{\infty} - \tau < 10$.  We use a fast and accurate analytic approximation of the second line to evaluate $\tilde Z$ when $\tau_{\infty}-\tau\ge 10$.  For example, in the lower troposphere,

\begin{equation}
{{\tilde Z} \over {2 \pi}}=-{{d{\tilde B}(\tau)}\over {d \tau}}  \Big[{4\over 3} - 2 E_4(\tau) \Big]\\.
\label{vn15}
\end{equation}

The spectral forcing, $\tilde F$, is defined as the difference between the spectral flux $\pi\tilde B_s$ through a transparent atmosphere from a black surface with temperature $T_s$, and the spectral flux $\tilde Z$ for an atmosphere with greenhouse gases,

\begin{equation}
\tilde F=\pi \tilde B_s-\tilde Z\\.
\label{vn58}
\end{equation}

\noindent The frequency integrals of the flux (\ref{b5}) and the forcing (\ref{vn58}) are

\begin{eqnarray}
Z&=&\int_0^{\infty}d\nu\,\tilde Z,\label{b58}\\
F&=&\int_0^{\infty}d\nu\,\tilde F=\sigma_{\rm SB}T_0^4-Z,\label{b60}\\
\end{eqnarray}

\noindent where $\sigma_{SB}$ is the Stefan Boltzmann constant.

\section{Intensity and Flux\label{rgm}}
In this section we discuss model atmospheres with greenhouse gas concentrations comparable to those of the year 2020.  The spectral flux has a complicated dependence on frequency $\nu$, the altitude profiles of the temperature $T$ and greenhouse gas concentrations $C^{\{i\}}$, and latitude.

For the standard atmosphere, the optical depths of (\ref{b11}) can be extremely large i.e. $\tau_{\infty}>>1$ at so called ``blanket" frequencies $\nu$ near the centers of the absorption lines.  For blanket frequencies a photon emitted near the surface is unlikely to escape the Earth's atmosphere because it has a high probability of being reabsorbed by a greenhouse gas molecule.  The extreme opposite of a blanket frequency is a ``window" frequency, where there is little absorption, and $\tau_{\infty}\approx 0$.  For window frequencies most radiation reaching space comes from the surface, with minor contributions from greenhouse gases.

From the first line of (\ref{vn14}), we see that for frequencies where the atmosphere is optically thick, with $\tau_{\infty} >> 1$, the flux at the top of the atmosphere is very nearly $\tilde Z(\tau_{\infty}) = \int_0^{\tau_{\infty}} d\tau^{\prime} E_2(\tau_{\infty} - \tau^{\prime})\tilde B (\tau')$.  For an isothermal atmosphere, with constant brightness $\tilde B$, half of the flux at the top of the atmosphere will come from altitudes above an emission optical depth $\tau_e$, defined by 

\begin{equation}
\int_{\tau_e}^{\tau_{\infty}} d\tau' E_2(\tau_{\infty}-\tau') = {1\over 2} \int_0^{\tau_{\infty}} d\tau' E_2(\tau_{\infty} - \tau').
\label{am0}
\end{equation}

\noindent Recalling that $E_2(\tau) = - dE_3(\tau)/d\tau$ and that $E_3(0) = 1/2$, we see that (\ref{am0}) implies that the value of the emission optical depth $\tau_e$ is given by

\begin{equation}
E_3(\tau_{\infty} - \tau_e) = 1/4, {\rm\ or\ } \tau_{\infty} - \tau_e = 0.41904.
\end{equation}

\begin{figure}[t]
\includegraphics[height=100mm,width=.95\columnwidth]{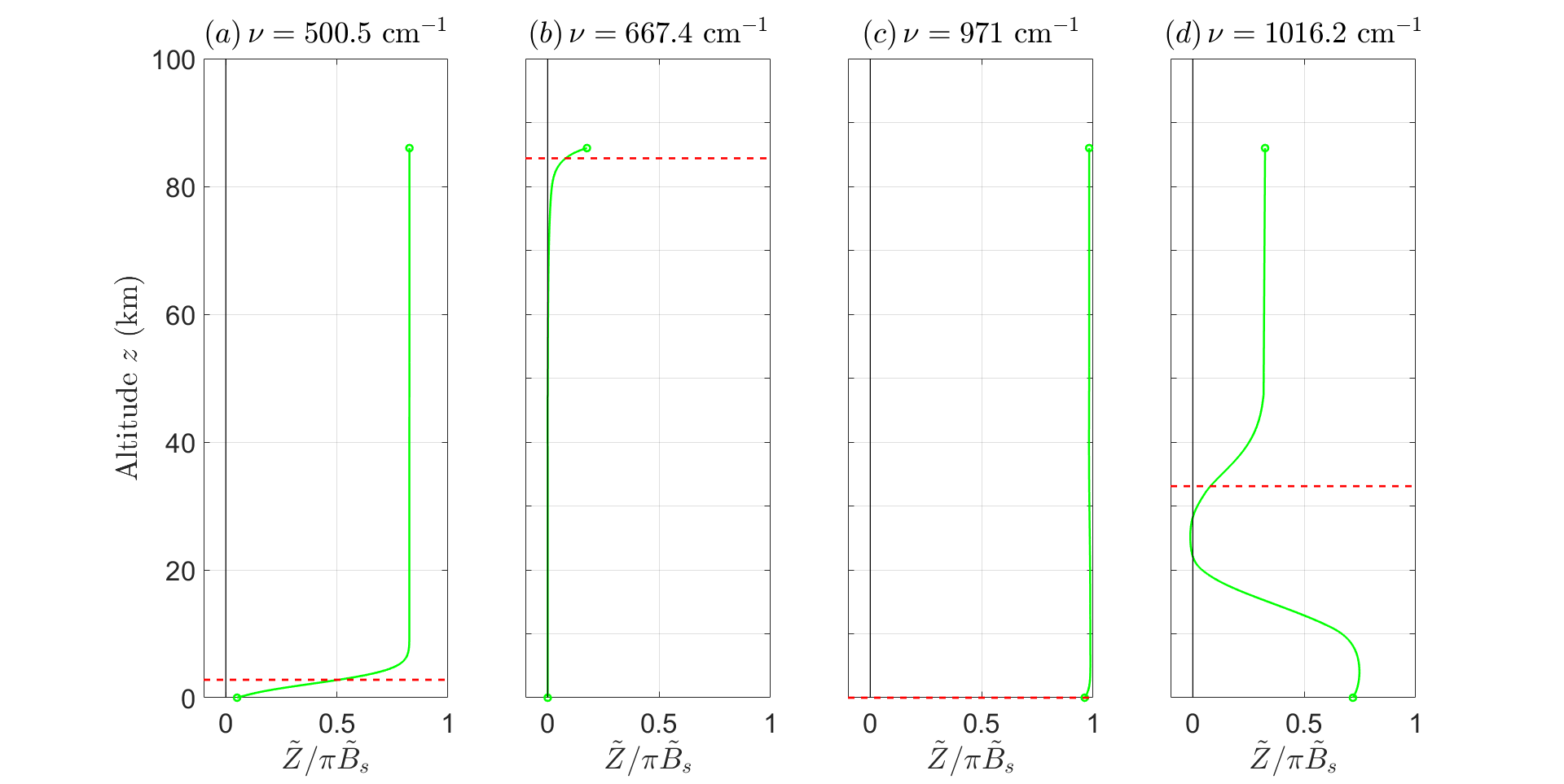}
\caption{Spectral fluxes $\tilde Z$ for representative
frequencies, $\nu$ indicated:  (a) $\tau_{\infty} = 2.82,\, z_e = 2.80$ km,\ $\tilde B_s= 134.2 \hbox{  i.u.}$; (b) $\tau_{\infty}=51688,\, z_e = 84.8$ km,\ $\tilde B_s = 131.8 \hbox{  i.u.}$; (c) $\tau_{\infty}=0.029,\, z_e = 0$ km,\ $\tilde B_s = 86.4 \hbox{  i.u.}$; (d) $\tau_{\infty}=7.54,\, z_e = 33.0$ km,\ $\tilde B_s = 79.3 \hbox{  i.u.}$. Here the spectral intensity unit is 1 i.u. = 1  mW m$^{-2}$ cm sr$^{-1}$.  \label{tZ500-1016}}
\end{figure}

Fig. \ref{tZ500-1016} shows a plot of upwards flux as a function of altitude at four different frequencies.  Fig. \ref{tZ500-1016}(a) shows a moderate blanket frequency $\nu = 500.5$ cm$^{-1}$, where the optical depth, $\tau_{\infty} =2.82$.  The emission height $z_e= 2.80$ km is in the lower troposphere.  From inspection of Fig. \ref{LI}, we see that transitions of the pure rotation spectrum of H$_2$O dominate the atmospheric opacity at the frequency, $\nu = 500.5$ cm$^{-1}$.

Fig. \ref{tZ500-1016}(b) shows an extreme blanket frequency $\nu = 667.4$ cm$^{-1}$ coinciding with a peak in the CO$_2$ absorption cross section where the optical depth is $\tau_{\infty}= 51,688$, and where the emission height is $z_e=84.8$ km, just below the mesopause.

Fig. \ref{tZ500-1016}(c) shows an extreme window frequency, $\nu = 971$ cm$^{-1}$, where the optical depth is only $\tau_{\infty} = 0.029.$  At this frequency, and in the absence of clouds, surface radiation reaches space with negligible attenuation by greenhouse gases.  The band centered on the wavelength $\lambda = 1/(971 \hbox{ cm}^{-1}) =10.3\, \mu{\rm m}$, is therefore called the ``clean infrared window."

Fig. \ref{tZ500-1016}(d) shows a blanket frequency, $\nu = 1016.2$ cm$^{-1}$, with a moderate optical depth $\tau_{\infty} = 7.54$
in the O$_3$ band. Not surprisingly, the emission height, $z_e = 33.0$ km, is in the upper stratosphere, where Fig. \ref{GGNT} shows
that the O$_3$ concentration is maximum.

\begin{figure}[t]
\includegraphics[height=100mm,width=1.0\columnwidth]{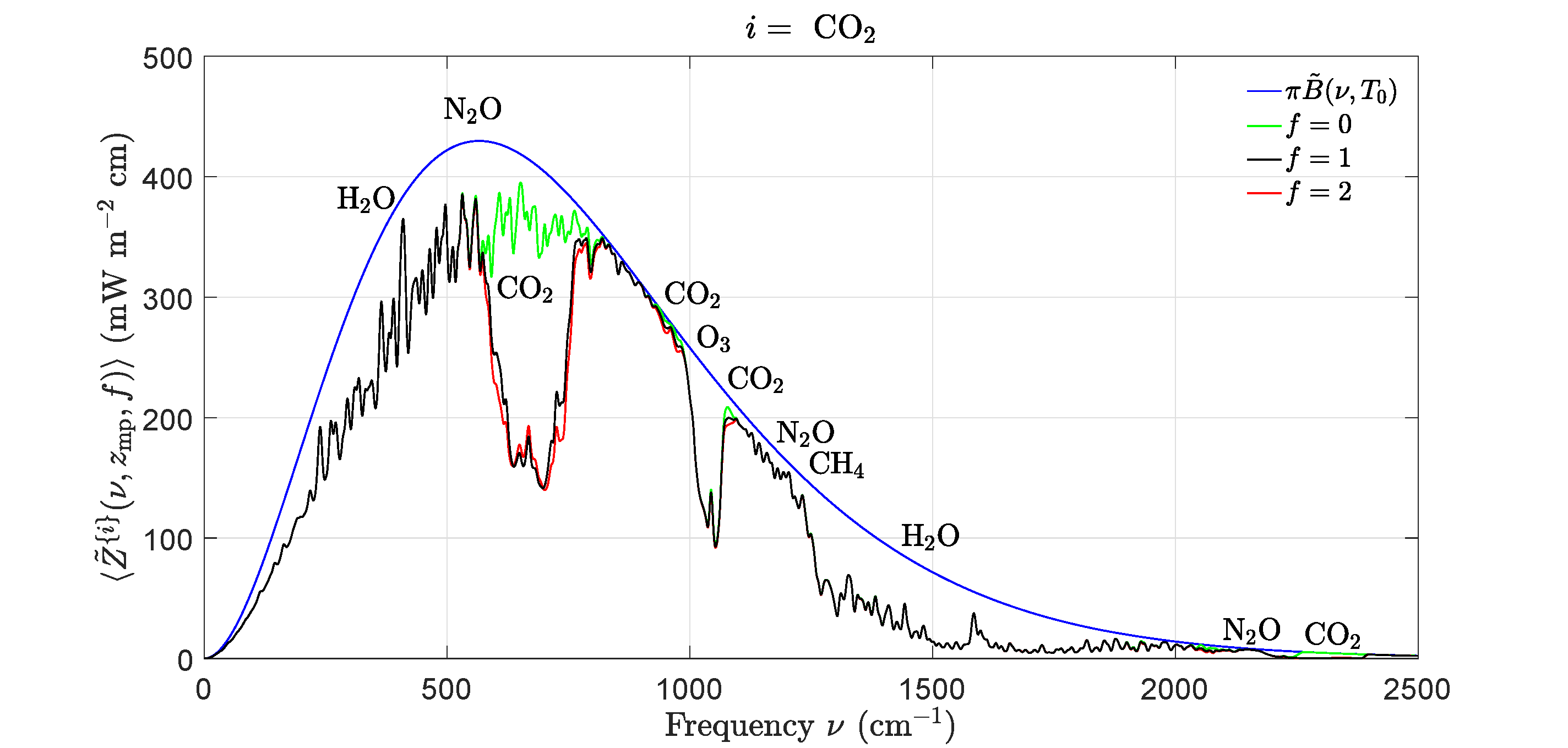} 
\caption{Effects of changing concentrations of carbon dioxide, CO$_2$ on the filtered spectral flux $\langle \tilde Z^{\{i\}}(\nu,z_{\rm mp},f)\rangle $ of (\ref{am10}) at the mesopause altitude, $z_{\rm mp}=86$ km.  The  width of the filter (\ref{am6}) was $\Delta\nu = 3$ cm$^{-1}$. The smooth blue line is the spectral flux, $\tilde Z =\pi\tilde B(\nu,T_0)$ from a surface at the temperature $T_0=288.7$ K for a transparent atmosphere with no greenhouse gases.  The green line is $\langle \tilde Z^{\{i\}}(\nu,z_{\rm mp},0)\rangle $ with the CO$_2$ removed but with all the other greenhouse gases at their standard concentrations.  The black line is
$\langle \tilde Z^{\{i\}}(\nu,z_{\rm mp},1)\rangle$ with all greenhouse gases at their standard concentrations. The red line is $\langle \tilde Z^{\{i\}}(\nu,z_{\rm mp},2)\rangle $ for twice the standard concentration of CO$_2$ but with all the other greenhouse gases at their standard concentrations.
Doubling the standard concentration of CO$_2$ (from 400 to 800 ppm) would cause a forcing increase (the area between the black and red lines)  of $\Delta F^{\{i\}} =3.0$ W m$^{-2}$, as shown in Table \ref{int}.
\label{CO2}}
\end{figure}

\begin{figure}[t] 
\includegraphics[height=100mm,width=1.0\columnwidth]{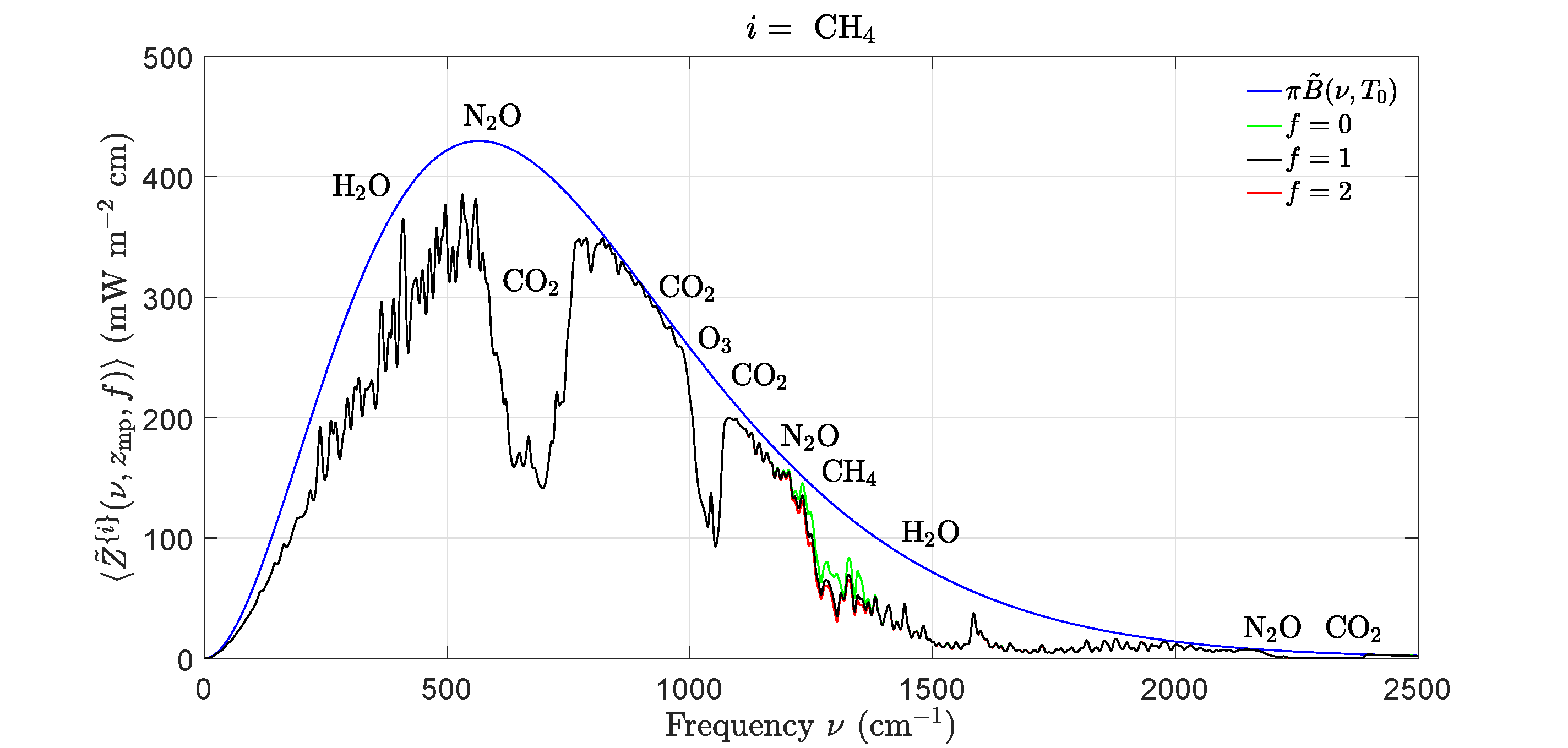}     
\caption{Effects of changing concentrations of methane, CH$_4$, on the filtered spectral flux $\langle \tilde Z^{\{i\}}(\nu,z_{\rm mp},f)\rangle $ of (\ref{am10}) at the mesopause altitude, $z_{\rm mp}=86$ km.  The blue and black lines have the same meanings as for Fig. \ref{CO2}. The green line is $\langle \tilde Z^{\{i\}}(\nu,z_{\rm mp},0)\rangle $ with the CH$_4$ removed but with all the other greenhouse gases at their standard concentrations.  The red line is $\langle \tilde Z^{\{i\}}(\nu,z_{\rm mp},2)\rangle $ with twice the standard concentration of  CH$_4$ but with all the other greenhouse gases at their standard concentrations.
Doubling the standard concentration of CH$_4$ would cause a forcing increase (the area between the black and red lines)  of $\Delta F^{\{i\}} =0.7$ W m$^{-2}$, as shown in Table \ref{int}.
\label{CH4}}
\end{figure}

High resolution spectrometers on satellites seldom provide measurements of intensity $\tilde I$ with resolutions less than 1 cm$^{-1}$.  For comparison of modeled spectral intensity or flux, it is useful to plot filtered spectral quantities.

\begin{equation}	
\langle \tilde X\rangle(z,\nu) = \int_0^{\infty}d\nu\,'J(\nu,\nu\,')\tilde X(z,\nu\,').
\label{am2}
\end{equation}

\noindent The filter function $J(\nu,\nu\,')$ smooths out sharp changes with frequency.  It is normalized so that

\begin{equation}	
\int_{-\infty}^{\infty}d\nu J(\nu,\nu\,')=1.
\label{am4}
\end{equation}

\noindent From (\ref{am2}) and (\ref{am4}) we see that the unfiltered spectral flux $\tilde Z$ and filtered spectral flux $\langle \tilde Z\rangle$ have the same frequency integral

\begin{equation}	
Z=\int_{0}^{\infty}d\nu \tilde Z=\int_{0}^{\infty}d\nu \,\langle\tilde Z\rangle,
\label{am8}
\end{equation}

\noindent and represent the same total flux $Z$.  We found it convenient to use a Gaussian filter function with a width parameter $\Delta \nu$,

\begin{equation}	
J(\nu,\nu\,')=\frac{e^{-(\nu-\nu\,')^2/2\Delta\nu^2}}{\sqrt{2\pi}\Delta\nu}.
\label{am6}
\end{equation}

\begin{figure}[t]                                                     \includegraphics[height=100mm,width=1.0\columnwidth]{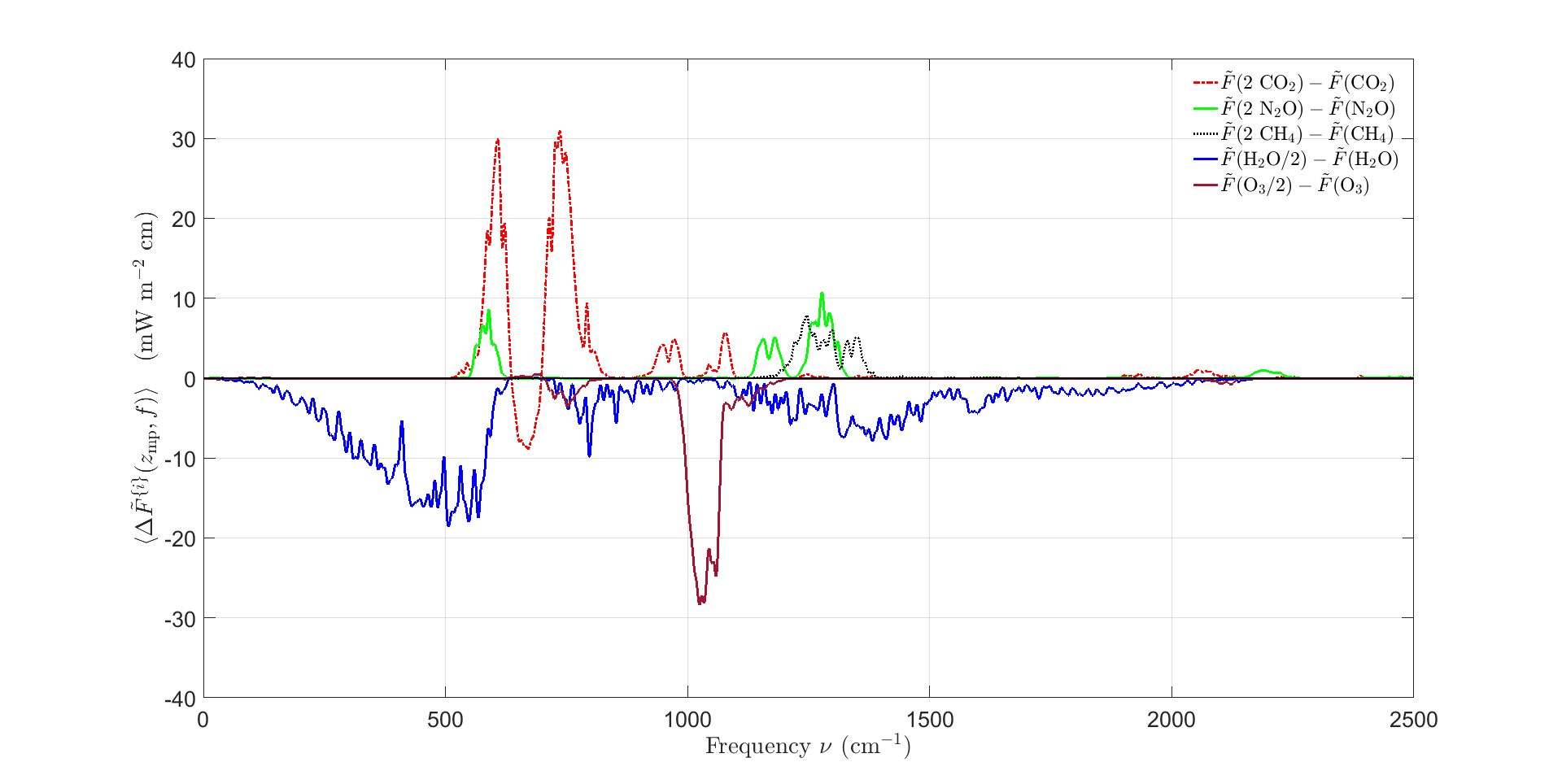} 
\caption{Spectral forcing increments (\ref{am12}) for doubled concentrations of CO$_2$, N$_2$O and CH$_4$. These are the magnified differences between the black and red curves of Figs. \ref{CO2} and \ref{CH4}. For most frequencies $\langle \Delta \tilde F^{\{i\}}(z_{\rm mp},2)\rangle$ is positive.  This is because doubling the concentrations of greenhouse gases shifts the emission heights $z_e$ of (\ref{am0}) to higher, colder altitudes in the troposphere.  An exception is the band of frequencies near the center of the exceptionally strong bending-mode band of CO$_2$ at 667 cm$^{-1}$.  Here doubling CO$_2$ moves the emission heights to higher, warmer altitudes of the stratosphere, where molecules can more efficiently radiate heat to space.  Also shown are forcing increments for halved concentrations of H$_2$O and O$_3$. Halving ensures that the relative humidity does not exceed 100\%, and reduces the clutter of the graph.
\label{DFGG}}
\end{figure}

The effects on radiative transfer of changing the column density of the $i$th greenhouse gas to some multiple $f$ of the standard value, $\hat N^{\{i\}}_{\rm sd}$, can be displayed with filtered spectral fluxes

\begin{equation}
\langle\tilde Z^{\{i\}}(\nu,z,f)\rangle=\langle\tilde Z(\nu, z,\hat N^{\{1\}}_{\rm sd},\ldots,\hat N^{\{i-1\}}_{\rm sd},f\hat N^{\{i\}}_{\rm sd}, \hat N^{\{i+1\}}_{\rm sd},\ldots,
\hat N^{\{n\}}_{\rm sd})\rangle.
\label{am10}
\end{equation}

\noindent Figs. \ref{CO2} and \ref{CH4} show how varying the concentrations of CO$_2$ and CH$_4$ affect the filtered spectral fluxes at the mesopause altitude, $z_{\rm mp} = 86$ km.  Expanded views of the differences between the flux for standard and doubled concentrations of greenhouse gases are shown in Fig. \ref{DFGG}, where we display

\begin{equation}
\langle\Delta \tilde F^{\{i\}}(z_{\rm mp},2)\rangle=\langle\tilde Z^{\{i\}}(\nu,z_{\rm mp},1)\rangle-\langle\tilde Z^{\{i\}}(\nu,z_{\rm mp},2)\rangle.
\label{am12}
\end{equation}

\begin{figure}[t]   
\includegraphics[height=100mm,width=1.0\columnwidth]{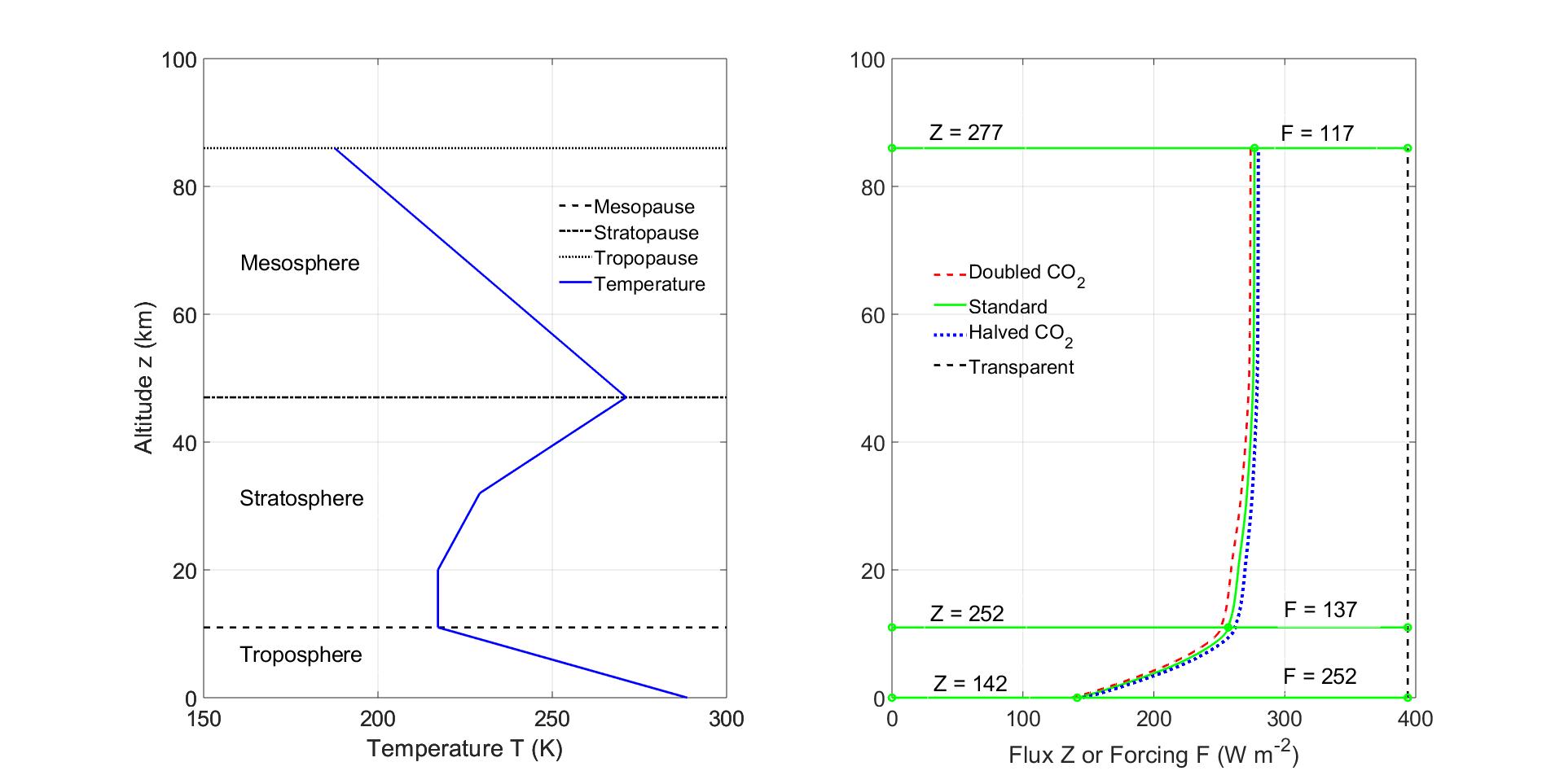} 
\caption{{\bf Left.} Midlatitude standard temperature profile. {\bf Right.} Altitude dependence of frequency integrated flux $Z$ of (\ref{b58}).  The flux for three concentrations of CO$_2$ are shown, the standard concentration, $C^{\{i\}}_{\rm sd}= 400$ ppm of Fig. \ref{GGNT},  twice and  half that value. The other greenhouse gases have the standard concentrations of Fig. \ref{GGNT}. The vertical dashed line is the flux $\sigma_{\rm SB}T_0^4=394$ W m$^{-2}$ for a transparent atmosphere with a surface temperature $T_0 = 288.7$ K. The forcings $F_s$ that follow from (\ref{b60}) at 0 km, 11 km and 86 km are 252, 137 and 117 W m$^{-2}$ respectively.
\label{ZzCM}}
\end{figure}

Integrating spectral fluxes, $\tilde Z$, like those of Fig. \ref{CO2}, over all frequencies in accordance with (\ref{b58}) gives $Z$, the frequency integrated flux shown in the right panel of Fig. \ref{ZzCM}. The calculations used the temperature profile of Fig. \ref{GGNT}, which is shown in the left panel of Fig. \ref{ZzCM}. A doubling of CO$_2$ concentration results in a 3 W/m$^2$ decrease in the top of the atmosphere flux.  This positive forcing changes the temperature profile as is discussed in Section 7.  

\begin{figure}[t]
\includegraphics[height=100mm,width=1.0\columnwidth]{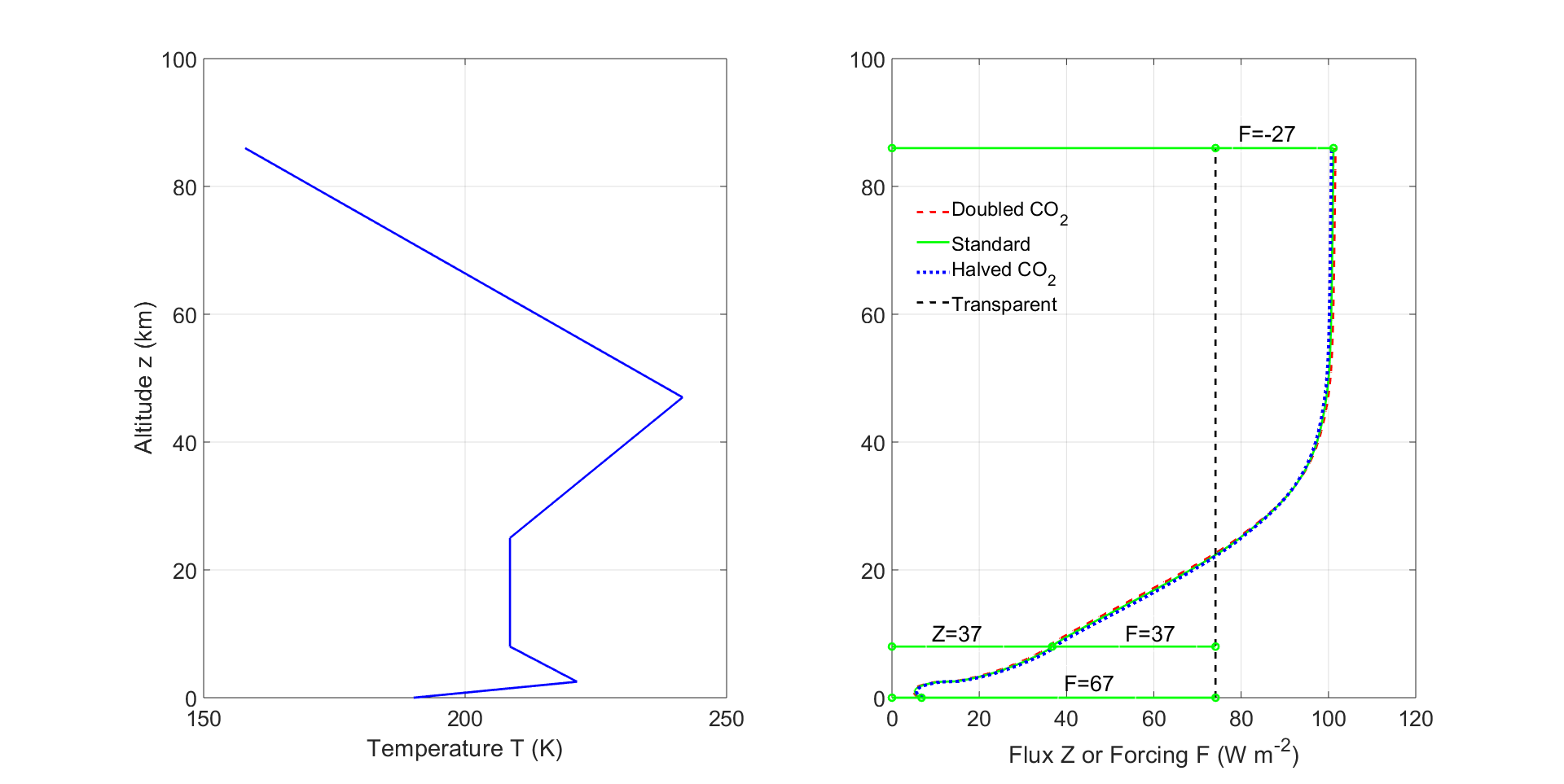} 	
\caption{Quantities analogous to those of Fig. \ref{ZzCM} at the South Pole.  There is a strong temperature inversion at an altitude of 2.5 km above the ice surface.  The flux for three concentrations of CO$_2$ are shown, the standard concentration, $C^{\{i\}}_{\rm sd}= 400$ ppm of Fig. \ref{GGNT}, twice and half that value.  The other greenhouse gases have the standard concentrations of Fig. \ref{GGNT}.  The relatively warm greenhouse-gas molecules in the atmosphere above the cold surface cause the Earth to radiate more heat to space from the poles than it could without greenhouse gases \cite{Smithuesen}.
\label{ZzCA}}
\end{figure}%

Thermal radiative fluxes depend on latitude.  They are larger near the equator where the surface is relatively warm than near the poles, where the surface is colder and where wintertime temperature inversions often form in the lower troposphere.  For example, Fig. \ref{ZzCA} is the analog of Fig. \ref{ZzCM} for Antarctica.  For Fig. \ref{ZzCA} we used a five segment temperature profile with altitude breakpoints at $ \zeta=[0, 2.5, 8, 25, 47, 86]$ km. The low tropopause at 8 km and the strong, wintertime temperature inversion, peaking at 2.5 km, are both characteristic of the nighttime poles. The lapse rates between the break points were $L = [-12.5, 2.3, 0,-1.5, 2.1]$ K km$^{-1}$.  The surface temperature in Antarctica was taken to be $T_0 = 190$ K and the surface pressure was set to be $p_0= 677$ hPa, because of the high elevation of the ice surface, about $2.7$ km above mean sea level.  Doubling CO$_2$ causes the negative forcing in the Antarctic opposite that shown in Fig. \ref{GGNT}.  This is due to the temperature inversion which means an increase in CO$_2$ causes more infrared radiation to escape to space, creating a negative greenhouse effect \cite{Smithuesen}. 

\begin{figure}[h]
\includegraphics[height=100mm,width=1.0\columnwidth]{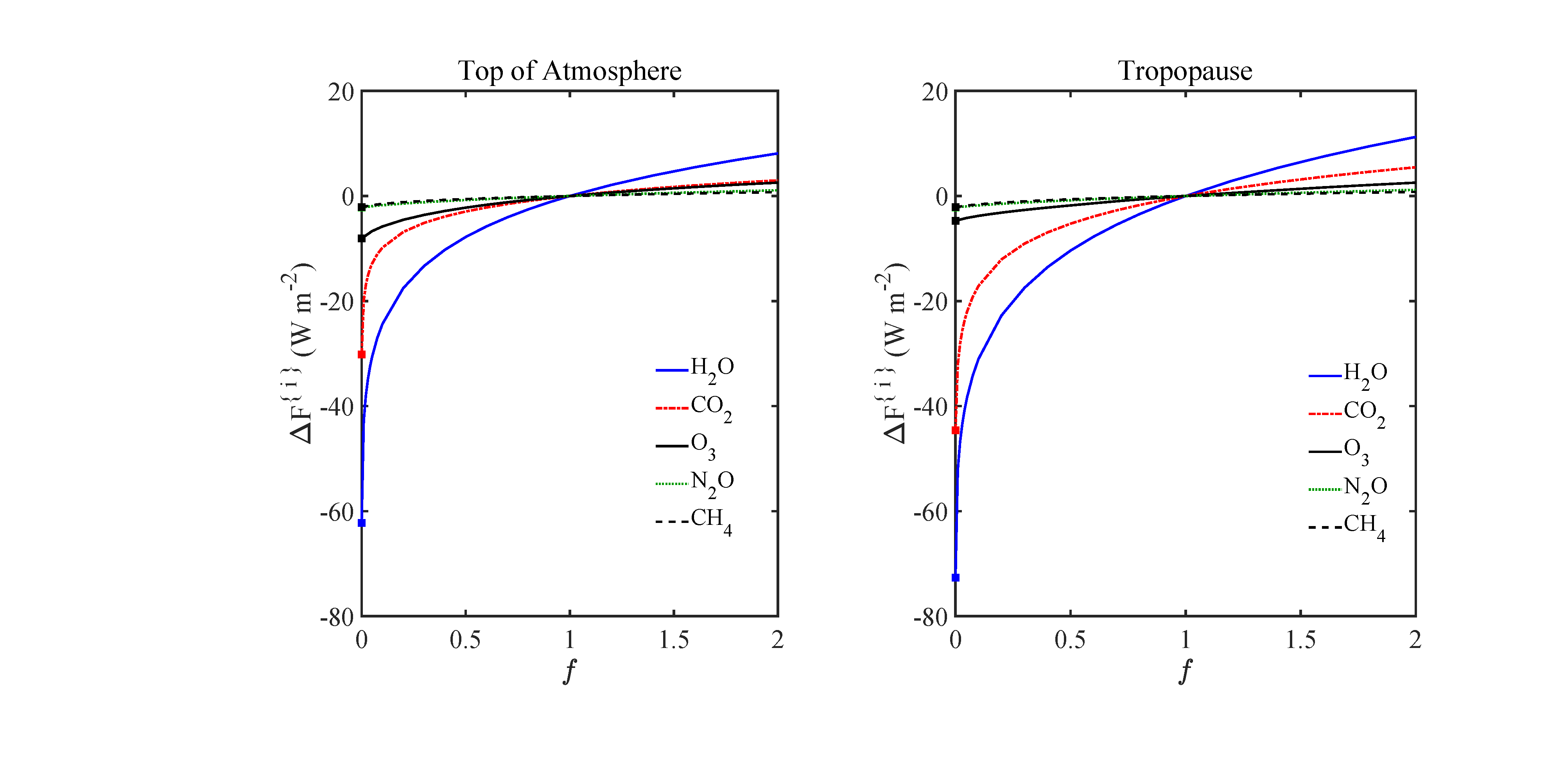} 
\caption{Dependence of partial forcing increments $\Delta F^{\{i\}}$ of (\ref{cdp18}) on greenhouse gas multiplicative factor, $f=N^{\{i\}}/N^{\{i\}}_{\rm sd}$.  At the standard column densities, with $f=1$, the incremental forcings are well into the saturation regime, with $d\Delta F^{\{i\}}(1)/df< d\Delta F^{\{i\}}(0)/df $ for all 5 gases.  For the most abundant greenhouse gases, H$_2$O and CO$_2$, the saturation effects are extreme, with per-molecule forcing powers suppressed by four orders of magnitude at standard concentrations $(f=1)$ with respect to the low-concentration, optically thin limit $(f=0)$. For CO$_2$, N$_2$O, and CH$_4$, the areas bounded by the green and black curves of Figs. \ref{CO2} and \ref{CH4} give the values, $-\Delta F^{\{i\}}$ for $f=0$, and the areas bounded by the black and red curves give $\Delta F^{\{i\}}$ for $f=2$, as discussed for Fig. \ref{DFGG}.  See the text and Table \ref{int} for more details. \label{DFDC}}
\end{figure}

\section{Concentration Dependence of Forcing\label{cdp}}

\begin{table}
\begin{center}
\begin{tabular}{||l| r r|r r|r r|r r|}
\hline
&\multicolumn{2}{c|}{$F^{\{i\}}_{\rm sd}(z)$}  &\multicolumn{2}{c|}{$\Delta F^{\{i\}}(z,0)$}&\multicolumn{2}{c|}{$\Delta F^{\{i\}}(z,1/2)$}&\multicolumn{2}{c|}{$\Delta F^{\{i\}}(z,2)$} \\
\hline
$i\setminus z$ & $z_{\rm tp}$&$z_{\rm mp}$&$z_{\rm tp}$ & $z_{\rm mp}$&$z_{\rm tp}$&$z_{\rm mp}$&$z_{\rm tp}$&$z_{\rm mp}$ \\ [0.5ex]
\hline\hline

H$_2$O&81.6&71.6&-72.6&-62.2&-10.4&-7.8&11.2&8.1\\
\hline
CO$_2$&52.4&38.9&-44.6&-30.2&-5.3&-3.0&5.5&3.0\\
\hline
O$_3$&6.1&10.5&-4.7&-8.1&-1.8&-2.2&2.5&2.5\\
\hline
N$_2$O&4.4&4.7&-2.2&-2.2&-0.8&-0.8&1.2&1.1\\
\hline
CH$_4$&4.2&4.4&-2.1&-2.1&-0.6&-0.6&0.8&0.7\\
\hline\hline
$\sum_i$&148.7&130.1&-126.2&-104.8&&&&\\
\hline\hline
$F_{\rm sd}(z)$&137&117&137&117&&&&\\
\hline
\end{tabular}
\end{center}

\caption{Partial forcings $F^{\{i\}}_{\rm sd}(z)$ of (\ref{cdp12}) and partial forcing increments $\Delta F^{\{i\}}(z,f)$
of (\ref{cdp18}), all in units of W m$^{-2}$, at the altitudes $z_{\rm tp}= 11$ km of the tropopause and $z_{\rm mp}= 86$ km of the mesopause. The last row contains the forcings $F_{\rm sd}(z)$ of (\ref{cdp10}), shown in Fig. \ref{ZzCM},
when all greenhouse molecules are present simultaneously at their standard column densities $\hat N^{\{i\}}_{\rm sd}$.
Because of the overlapping absorption bands, $\sum_i F^{\{i\}}_{\rm sd}(z)>F_{\rm sd}(z)$, and $-\sum_i \Delta F^{\{i\}}(z,0)<F_{\rm sd}(z)$. \label{int}}
\end{table}

The frequency integrated forcing, $F$, of (\ref{b60}) depends on the altitude $z$ and on the
column densities of the five greenhouse gases given in Table \ref{acd}.
\begin{equation}
F=F(z,\hat N^{\{1\}},\ldots,\hat N^{\{5\}}).
\label{cdp8}
\end{equation}
We assume the temperature $T$ and densities $N^{\{i\}}$ have the same altitude profiles as in the midlatitude example of Fig. \ref{GGNT}.  An important special case of (\ref{cdp8}) is the forcing, $F_{\rm sd}$, when each greenhouse gas $i$ is present at its standard column density $\hat N^{\{i\}}_{\rm sd}$ of Table \ref{acd},
\begin{equation}
F_{\rm sd}(z)=F(z,\hat N^{\{1\}}_{\rm sd},\ldots,\hat N^{\{5\}}_{\rm sd}).
\label{cdp10}
\end{equation}
A second special case of (\ref{cdp8}) is the hypothetical, per molecule standard forcing, $F^{\{i\}}_{\rm sd}$, when the atmosphere contains only molecules of type $i$ at their standard column density, $\hat N^{\{i\}}=\hat N^{\{i\}}_{\rm sd}$, and the concentrations of the other greenhouse vanish, $\hat N^{\{j\}}=0$ if $j\ne i$,
\begin{equation}
F^{\{i\}}_{\rm sd}(z)=F(z,0,\ldots,0,\hat N^{\{i\}}_{\rm sd}, 0,\ldots,0).
\label{cdp12}
\end{equation}

We define the forcing power per added molecule as
\begin{equation}
P^{\{i\}}(z,\hat N^{\{1\}},\ldots,\hat N^{\{n\}})=\frac{\partial F}{\partial \hat N^{\{i\}}}.
\label{cdp14}
\end{equation}
The densities of greenhouse gases $j$ with $j\ne i$ are held constant in the
partial derivative of (\ref{cdp14}).  If the units of $F$ are taken to be W m$^{-2}$ and the units of $\hat N^{\{i\}}$ are taken to be molecules m$^{-2}$, then the units of $P^{\{i\}}$ will be W molecule$^{-1}$.

We define a finite forcing increment for the $i$th type of greenhouse molecule as
\begin{equation}
\Delta F^{\{i\}}(z,f)=F(z,\hat N^{\{1\}}_{\rm sd},\ldots,\hat N^{\{i-1\}}_{\rm sd},f\hat N^{\{i\}}_{\rm sd}, \hat N^{\{i+1\}}_{\rm sd},\ldots,
\hat N^{\{n\}}_{\rm sd})-F_{\rm sd}.
\label{cdp18}
\end{equation}
Differentiating (\ref{cdp18}) with respect to $f$ we find
\begin{equation}
\frac{\partial \Delta F^{\{i\}}}{\partial f}(z,f)=\hat N^{\{i\}}_{\rm sd}P^{\{i\}}_{\rm sd}(z,f),
\label{cdp20}
\end{equation}
where $P^{\{i\}}_{\rm sd}(z,f)$ is the forcing power per additional molecule of type $i$ when these molecules have the column density $\hat N^{\{i\}}=f\hat N^{\{i\}}_{\rm sd}$ and all other types of greenhouse molecules have their standard column densities.

The forcing increments (\ref{cdp18}) for the five greenhouse gases considered in this paper are shown as a function of $f$ in Fig. \ref{DFDC}. Forcing increments are also tabulated at representative altitudes $z$ and multiplicative factors $f$ in Table \ref{int}.  At both the top of the atmosphere and at the tropopause, we see that the forcing increment (\ref{cdp18}) is largest for abundant water molecules, H$_2$O, and is relatively small for the much more dilute greenhouse gases CH$_4$ and N$_2$O.  The incremental forcings are all in the saturation regime, with $\partial \Delta F^{\{i\}}/\partial f$ diminishing with increasing $f$.

In Table \ref{int}, the forcing decrements from removing H$_2$O, CO$_2$, O$_3$, N$_2$O and CH$_4$, $-62.2$, $-30.2$, $-8.1$, $-2.2$ and $-2.1$  W m$^{-2}$, are reasonably close to those calculated by Zhong and Haigh \cite{Zhong2013}. In their Table 1 they cite forcing decrements at the top of the atmosphere of $-70.6$, $-25.5$, $-7.0$, $-1.8$ and $-1.7$ W m$^{-2}$.  Zhong and Haigh seem to have taken the concentrations of N$_2$O and CH$_4$ to be independent of altitude. The altitude dependence of Fig. $\ref{GGNT}$ were used in our calculations.

Note from Table \ref{int} that doubling or halving the column density of CO$_2$ changes the forcing $F$ by almost the same amount, either at the tropopause or at the mesopause.  This dependence of forcing increments on the logarithm of the CO$_2$ column density was first pointed out by Arrhenius \cite{Arrhenius1908}.  Wilson and Gea-Banacloche \cite{Wilson12} explain how the approximate dependence of the CO$_2$ absorption cross section on frequency $\nu$, $\sigma^{\{i\}}=\sigma_e e^{-\lambda_e|\nu -\nu_e|}$, leads to the logarithmic forcing law.  Here $\sigma_e$ is the maximum cross section at the center frequency, $\nu_e= 667$ cm$^{-1}$, of the bending mode band.

\begin{table}
\begin{center}
\begin{tabular}{|c|c| r r| r  r|}
\hline
& &\multicolumn{4}{c|} {$\Delta F^{\{i\}}(z,f)  \hbox{ in  W m}^{-2}$ }\\
\hline
& &\multicolumn{2}{c|}{Ref. \cite {Collins2006} } &\multicolumn{2}{c|}{This Work}\\
\hline
$i$&$f$&$z_{\rm tp}$ &$z_{\rm mp}$ & $z_{\rm tp}$ & $z_{\rm mp}$\\
\hline\hline
H$_2$O&1.06 &1.4 &1.1&0.9&0.7\\ \hline
CO$_2$&2&5.5&2.8&5.5&3.0\\ \hline
O$_3$&1.1&&&0.3&0.3\\ \hline
N$_2$O&2&1.3&1.2&1.2&1.1\\ \hline
CH$_4$ &2&0.6&0.6&0.8&0.7\\
 \hline
\end{tabular}
\end{center}
\caption{Comparison of the forcing increments $\Delta F^{\{i\}}(z,f)$ of Collins {\it et al.}\cite{Collins2006} in column 3, and the results of Table \ref{int} and (\ref{cdp18}) in column 4, at the altitude $z_{\rm tp} = 11$ km of the tropopause and $z_{\rm mp} = 86$ km of the mesopause.  For H$_2$O, the relative increase, $f=1.06$, of the column density is approximately that caused by a 1 K increase of the surface temperature. \label{int2}}
\end{table}
The forcing increments in Table \ref{int2} are comparable to those calculated by others. For example, in column 3  we give the increments  $\Delta F^{\{i\}}(z,f)$ calculated by Collins et al \cite{Collins2006}, as
estimated from their Tables 2 and 8.  These are the results of averaging five separate line by line calculations.  In addition to line intensities, three of the calculations used a continuum CO$_2$ opacity, and
all five used a continuum H$_2$O opacity.  The physical origin of these continua is unclear.  They are added to make the calculations agree better with observations \cite{Edwards1992,Collins2006}. 
The forcings calculated in this paper, summarized in column 4, used only lines for the HITRAN data base and no continua.  Our values are fairly close to those of Collins et al.\cite{Collins2006}, with the largest discrepancy for H$_2$O.  The mesopause spectral intensities, calculated with only HITRAN lines and with no continuum contributions, are in excellent agreement with satellite measurements over the
Sahara Desert, the Mediterranean Sea and Antarctica, as discussed in Section \ref{ld}.

The three mesopause flux increments $\Delta F^{\{i\}}$ in  the fourth column of Table \ref{int2} for doubled concentrations of CO$_2$, N$_2$O and CH$_4$ sum to 4.8 W m$^{-2}$.  The calculated flux increment from simultaneously doubling CO$_2$, N$_2$O and CH$_4$ is the slightly smaller value, $\Delta F = 4.7$ W m$^{-2}$.
Similarly, the four mesopause flux increments $\Delta F^{\{i\}}$ in the fourth column of Table \ref{int2} for doubled concentrations of CO$_2$, N$_2$O and CH$_4$ as well as a factor of $f=1.06$ increase of H$_2$O concentration sum to 5.5 W m$^{-2}$.
The calculated flux increment from simultaneously doubling CO$_2$, N$_2$O and CH$_4$, and increasing the H$_2$O concentrations by a factor of $f=1.06$, is the slightly smaller value 5.3 W m$^{-2}$.  The ``whole" is  less than the sum of the parts, because of the interference of greenhouse gases that absorb the same infrared frequencies.

Table \ref{dPr} summarizes the forcing powers (\ref{cdp14}) per additional molecule in units of $10^{-22}$ W at
the tropopause altitude, $z_{\rm tp}=11$ km and at the  mesopause altitude, $z_{\rm mp}=86$ km. The surface temperature was $T_0=288.7$ K,
and the altitude profiles of temperature and number density were those of Fig. $\ref{GGNT}$. The first column lists the molecules we considered.  The numbers in the second column are forcing powers, $P^{\{i\}}_{\rm ot}(z)$, of (\ref{ot6}) in the optically thin limit. The numbers of the third column are forcing powers $P^{\{i\}}_{\rm sd}(z,0)$ from (\ref{cdp20}) for an atmosphere that previously had no molecules of type $i$ (so $\hat N^{\{i\}}=0$) but all other greenhouse molecules had standard concentrations, $\hat N^{\{j\}}=\hat N^{\{j\}}_{\rm sd}$ if $j\ne i$. The forcings of the third column are less than those of the second because of interference between absorption by different greenhouse gases. The numbers in the fourth column are the forcing powers $P^{\{i\}}_{\rm sd}(z,1)$ from (\ref{cdp20}) when a single molecule of type $i$ is added to an atmosphere that previously had standard  densities for all greenhouse gases, $\hat N^{\{j\}}=\hat N^{\{j\}}_{\rm sd}$. Saturation of the absorption suppresses the per-molecule forcing by about four orders of magnitude for the abundant greenhouse gases H$_2$O and CO$_2$. Saturation causes less drastic suppression of per-molecule forcings for the less abundant O$_3$, N$_2$O and CH$_4$.

\begin{table}
	\begin{center}
		\begin{tabular}{|l| c c | c c| c c |}
			\hline
			&\multicolumn{2}{c|}{$P^{\{i\}}_{\rm ot}(z)$}&\multicolumn{2}{c|}{$P^{\{i\}}_{\rm sd}(z,0)$}&\multicolumn{2}{c|}{$P^{\{i\}}_{\rm sd}(z,1)$}\\
			\hline
			$i\setminus z$& $z_{\rm tp}$ & $z_{\rm mp}$  & $z_{\rm tp}$ & $z_{\rm mp}$   & $z_{\rm tp}$  & $z_{\rm mp}$  \\ [0.5ex]
			\hline\hline
			H$_2$O&1.49&1.49&1.16&1.19& $3.3\times 10^{-4}$ & $2.5\times 10^{-4}$ \\
			\hline
			CO$_2$&2.73&3.45&2.24&2.53&$9.0\times 10^{-4}$&$4.9\times 10^{-4}$\\
			\hline
			O$_3$&2.00&5.69&1.68&4.57&$3.3\times 10^{-1}$&$3.8\times 10^{-1}$\\
			\hline
			N$_2$O&1.68&2.24&0.73&0.91&$2.1\times 10^{-1}$& $2.0\times 10^{-1}$\\
			\hline
			CH$_4$&0.51&0.71&0.21&0.27&$2.8\times 10^{-2}$&$2.6\times 10^{-2}$\\
			\hline
		\end{tabular}
	\end{center}
	\caption{Forcing powers (\ref{cdp14}) per additional molecule in units of $10^{-22}$ W at the altitude $z_{\rm tp} = 11$ km of the tropopause and $z_{\rm mp} = 86$ km of the mesopause. The surface temperature was $T_0=288.7$ K,
		and the altitude profiles of temperature and number density were those of Fig. $\ref{GGNT}$.  $P^{\{i\}}_{\rm ot}(z)$ of (\ref{ot6}) is for the optically-thin limit.  $P^{\{i\}}_{\rm sd}(z,0)$ from (\ref{cdp20}) is for an atmosphere that previously had no molecules of type $i$ (so $\hat N^{\{i\}}=0$) but all other greenhouse molecules had standard concentrations.
		$P^{\{i\}}_{\rm sd}(z,1)$ from (\ref{cdp20}) is for a single molecule of type $i$  added to an atmosphere that previously had standard  densities for all greenhouse gases.\label{dPr}}
\end{table}

\begin{figure}[h]
\includegraphics[height=100mm,width=1.0\columnwidth]{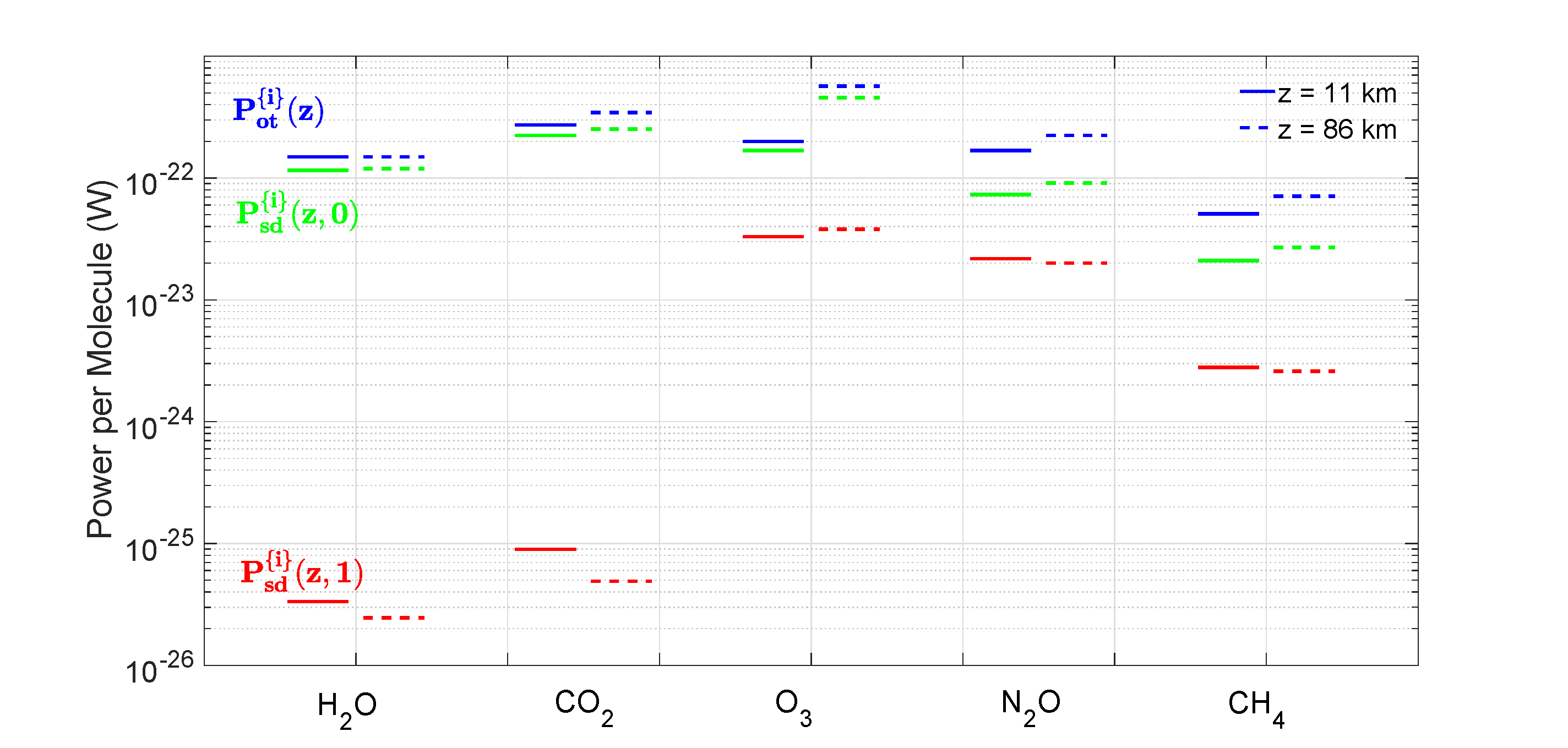} 
\caption{A graphical display of the per-molecule forcing powers of Table \ref{dPr}. At standard column densities the (red) powers, $P^{\{i\}}_{\rm sd}(z,1)$, for H$_2$O and CO$_2$ are suppressed by four orders of magnitude from their values in the optically thin limit (blue) where the powers are $P^{\{i\}}_{\rm ot}(z)$. This is due to strong saturation of the absorption bands. Saturation effects (difference between the blue and red lines) are much less for the minor gases, O$_3$, N$_2$O and CH$_4$. The green lines are the powers per molecule, $P^{\{i\}}_{\rm sd}(z,0)$, of the $i$th greenhouse gas in its low-concentration limit, but when the forcing power is suppressed by other gases at their standard densities.  Interference effects (difference between the blue and green lines) are more pronounced for N$_2$O and CH$_4$ than for H$_2$O and CO$_2$. Fig. \ref{LI} shows the strongest bands of O$_3$ overlap little with those of other greenhouse molecules, minimizing interference effects.
\label{Pbar}}
\end{figure}

We now consider the optically thin limit, where the concentrations of greenhouse gases are sufficiently low that the optical depths $\tau$ of (\ref{b10}) will be small, $\tau\ll 1$, for all frequencies $\nu$ and at all altitudes $z$.  The frequency integral of the spectral forcing (\ref{b58}) at altitude $z$ can then be written as

\begin{eqnarray}
F_{\rm ot}(z)&=&\sum_i\hat N^{\{i\}} P^{\{i\}}_{\rm ot}(z)
\label{ot2}
\end{eqnarray}

\noindent where the forcing power per greenhouse molecule of type $i$ is

\begin{equation}
P^{\{i\}}_{\rm ot}(z)=\frac{1}{2}\int_0^z dz'\frac{N^{\{i\}'}}{\hat N^{\{i\}}}\left[\Pi^{\{i\}}(T',T_0)-\Pi^{\{i\}}(T',T')\right]
+\frac{1}{2}\int_0^z dz'\frac{N^{\{i\}'}}{\hat N^{\{i\}}}\Pi^{\{i\}}(T',T').
\label{ot4}
\end{equation}

\noindent Here $N^{\{i\}'} = N^{\{i\}}(z')$,  $T'=T(z')$ and $T_0=T(0)$. 
The mean power absorbed by a greenhouse gas molecule of temperature $T$  from thermal equilibrium radiation of temperature $T'$ is

\begin{equation}
\Pi^{\{i\}}(T,T')= 4\pi\sum_{ul} S_{ul}^{\{i\}}(T)  \tilde B(\nu_{ul},T').
\label{ot6}
\end{equation}

\noindent For the special case of $T'=T$ we can substitute (\ref{lbl24}) into (\ref{ot6}) to find
\begin{eqnarray}
\Pi^{\{i\}}(T,T)&=&\sum_{ul}W_u^{\{i\}}(T)\Gamma_{ul}^{\{i\}}E_{ul}^{\{i\}}.
\label{ot8}
\end{eqnarray}
Since we are considering a single isotopologue, we have set $\eta_u=1$ in (\ref{lbl24}).  The three factors in the summed terms of (\ref{ot8}) are  the probability $W_u^{\{i\}}(T)$ to find the molecule in the upper state $u$, the radiative decay rate $\Gamma_{ul}^{\{i\}}$ from the upper level $u$ to the lower level $l$ and the mean energy $E_{ul}^{\{i\}}$ of the emitted photon.  This is obviously the total power radiated by a molecule of temperature $T$.  For a molecule of temperature $T$ in thermal equilibrium with radiation of the same temperature, the radiative power absorbed by the molecule is equal to the spontaneous radiative power it emits. The forcing powers per molecule are summarized graphically in Fig. \ref{Pbar}.

\section{Temperature and Forcing}

The forcings due to instantaneous changes of greenhouse gas concentrations can be calculated quite accurately.  Temperature changes induced by the forcings are less clearly defined because various feedbacks change the temperature profile of the atmosphere.  After doubling CO$_2$ concentrations, a new, steady state will eventually be established by these feedback processes.

\begin{figure}[t]
\includegraphics[height=80mm,width=0.9\columnwidth]{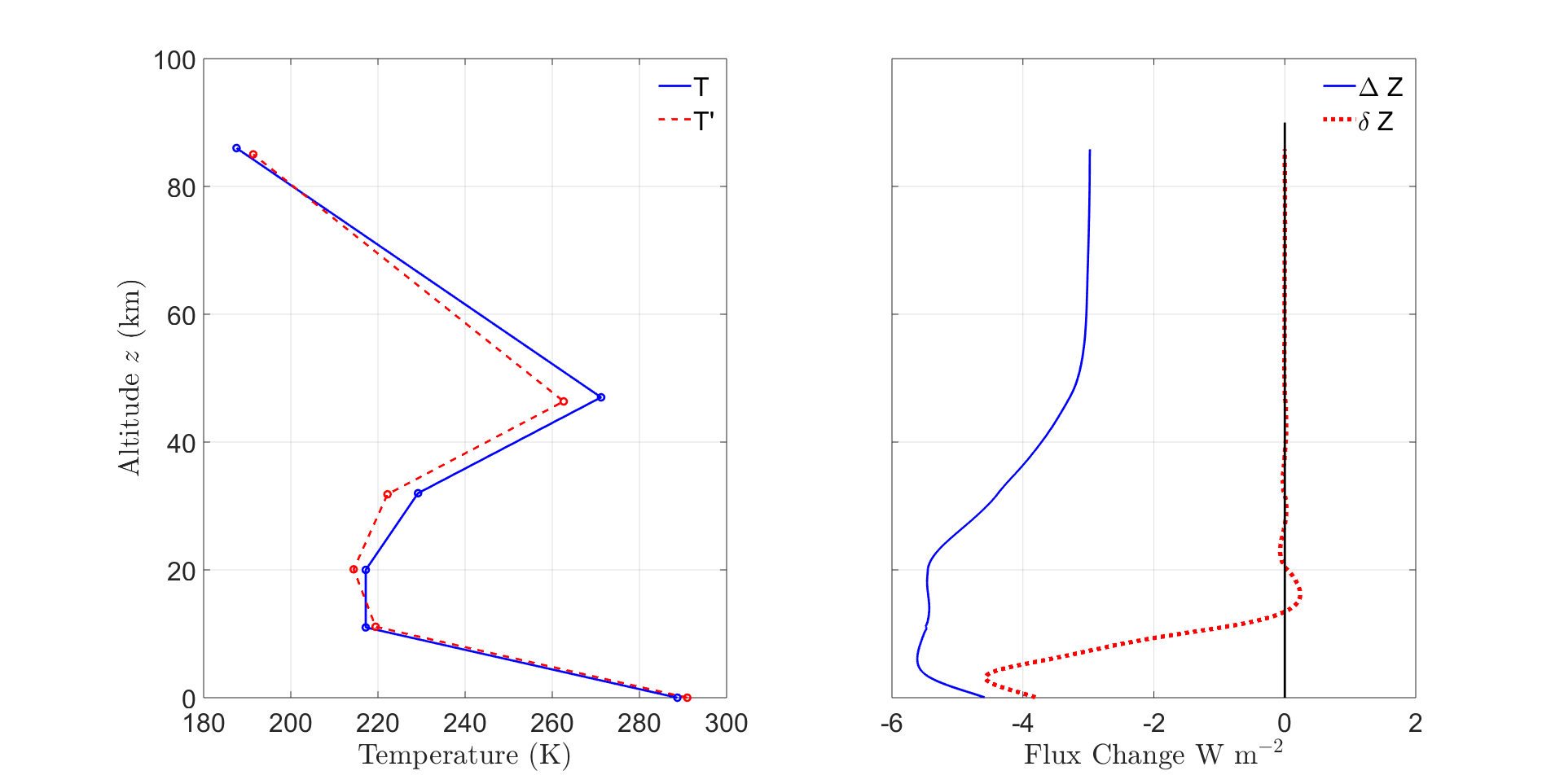} 
\caption{Temperature and Flux Adjustments.  
{\bf Left.} An initial temperature profile $T$ (continuous blue line), which is characteristic of mid latitudes, and the adjusted profile $T' = T + \Delta T$ (dotted red line), from (\ref{rce26}).  {\bf Right.}  The continuous blue line is the altitude profile of the instantaneous flux decrease $\Delta Z$ of (\ref{rce2}) caused by increasing CO$_2$ concentrations from $400$ ppm to $800$ ppm, with no change in the profiles $T$ of temperature or $C_w$ of water-vapor concentration.  The dotted red line shows the residual flux change $\delta Z$ of (\ref{rce8}) for the adjusted temperature profile $T'$ of the left panel, together with the adjusted water-vapor concentration $C_w'$ to keep relative humidity constant at all altitudes.  These adjustments restore the flux as nearly as possible to its original value for altitudes above the tropopause, the criterion for maintaining radiative-convective equilibrium.
\label{TempFluxAdjust}}
\end{figure}

As shown in Fig. \ref{ZzCM}, we have computed the upwards flux
$Z = Z(C_g, C_w, T)$ where $C_g$ and $C_w$ are the initial concentrations of some greenhouse gas and water vapor respectively, and $T$ is the temperature.  The flux depends on altitude $z$ because $C_g$, $C_w$ and $T$ are each functions of altitude.  
In the right panel of  Fig. \ref{ZzCM} we showed that changing the concentration $C_g$ of the greenhouse gas CO$_2$ to twice its value, $C_g'=2C_g$, while holding all other atmospheric properties the same,  changed the flux from its initial value $Z(C_g, C_w,T)$ to a slightly smaller value, $Z(C_g', C_w,T)$.  In this hypothetical  ``instantaneous" process there is no change in the atmospheric temperature $T=T(z)$.  The concentrations of all other greenhouse gases, most notably, the concentration $C_w = C_w(z)$ of water vapor, also remain the same.  The difference between the flux before and after addition of the greenhouse gas  is called the {\it instantaneous forcing increment}, and can be written as
\begin{equation}
\Delta F = -\Delta Z=Z(C_g, C_w, T) - Z(C_g', C_w,T).
\label{rce2}
\end{equation}
The flux increment $\Delta Z$ for doubling CO$_2$ concentrations is too small to be seen clearly in Fig. \ref{ZzCM} so we have plotted an expanded version  in the right panel of Fig. \ref{TempFluxAdjust}. The magnitude of the flux increment is somewhat greater in the lower atmosphere than at higher altitudes.

In steady state the atmosphere is commonly assumed to be in {\it radiative-convective equilibrium}, described in 1967 by Manabe and Wetherald\cite{Manabe1967}.  If the concentration $C_g$ of a greenhouse gas other than water vapor instantaneously changes to a new value $C_g'=C_g+\Delta C_g$, the atmosphere will no longer be in equilibrium.  To restore radiative-convective equilibrium the temperature profile will change to $T'=T+\Delta T$ and the water-vapor concentration will change to $C_w'=C_w+\Delta C_w$.

The first criterion for radiative-convective equilibrium is no change in thermal flux at the top of radiative atmosphere, which we take to be the mesopause altitude $z_{\rm mp}=\zeta_5$,
\begin{equation}
Z_{\rm mp}(C_g', C_w', T')= Z_{\rm mp}(C_g, C_w,T),\quad\hbox{for}\quad z=\zeta_5.
\label{rce4}
\end{equation}
This criterion assumes that net solar heating remains the same
and is balanced by thermal radiation to space.

Above the tropopause, vertical convection is negligible.  Thermal radiation carries off heat resulting from the absorption of solar ultraviolet radiation by ozone, and also any heating by horizontal convection.  In radiative equilibrium, the volume heating rate is balanced by the thermal-radiation cooling rate, $dZ/dz$.  So a second criterion for radiative-convective equilibrium is no change in either the heating or cooling rates
\begin{equation}
\frac{dZ(C_g', C_w', T')}{dz} =\frac{dZ(C_g, C_w, T)}{dz}\quad\hbox{for}\quad
\zeta_1<z<\zeta_5.
\label{rce6}
\end{equation}
We can integrate (\ref{rce6}) down in altitude from the top of the atmosphere, using (\ref{rce4}) as a boundary condition.  This gives a combined criterion for radiative equilibrium above the tropopause,
\begin{equation}
\delta Z = Z(C_g', C_w', T')- Z(C_g, C_w,T)=0\quad\hbox{for}\quad \zeta_1<z<\zeta_5.
\label{rce8}
\end{equation}

Below the tropopause, much of the heat is transported by vertical convection rather than radiation.  Manabe and Wetherald suggested that  for convective equilibrium in the troposphere, the temperature lapse rate, $-\partial T/\partial z$ should equal (or not exceed) an equilibrium value, $L$, that is a known function of the altitude $z$ and of the surface temperature $\theta_0$.  Formally, the criterion is
\begin{equation}
-\frac{\partial}{\partial z} T(z,\theta_0)=L(z,\theta_0)\quad\hbox{for}\quad z<\zeta_1.
\label{rce10}
\end{equation}
Integrating (\ref{rce10}) we see that the tropospheric temperature is a function of altitude $z$ and surface temperature $\theta_0$, given by
\begin{equation}
T(z,\theta_0)=\theta_0-\int_0^z dz' L(z',\theta_0).
\label{rce12}
\end{equation}

Tropospheric lapse rates measured by radiosondes are often quite complicated \cite{Radiosondes}.  Simplified functions are normally used to approximate lapse rates.  Manabe and Wetherald\cite{Manabe1967} made extensive use of the altitude-independent tropospheric lapse rate
\begin{equation}
\bar L= 6.5 \hbox{ K km}^{-1}.
\label{lr2}
\end{equation}
This can be thought of as the average, over the troposphere, of a more complicated lapse rate
\begin{equation}
\bar L=-\frac{1}{\zeta_1}\int_0^{\zeta_1} dz\frac{\partial T}{\partial z}  =\frac{\theta_0-\theta_1}{\zeta_1}.
\label{lr3}
\end{equation}

\begin{figure}[t]
\includegraphics[height=80mm,width=0.9\columnwidth]{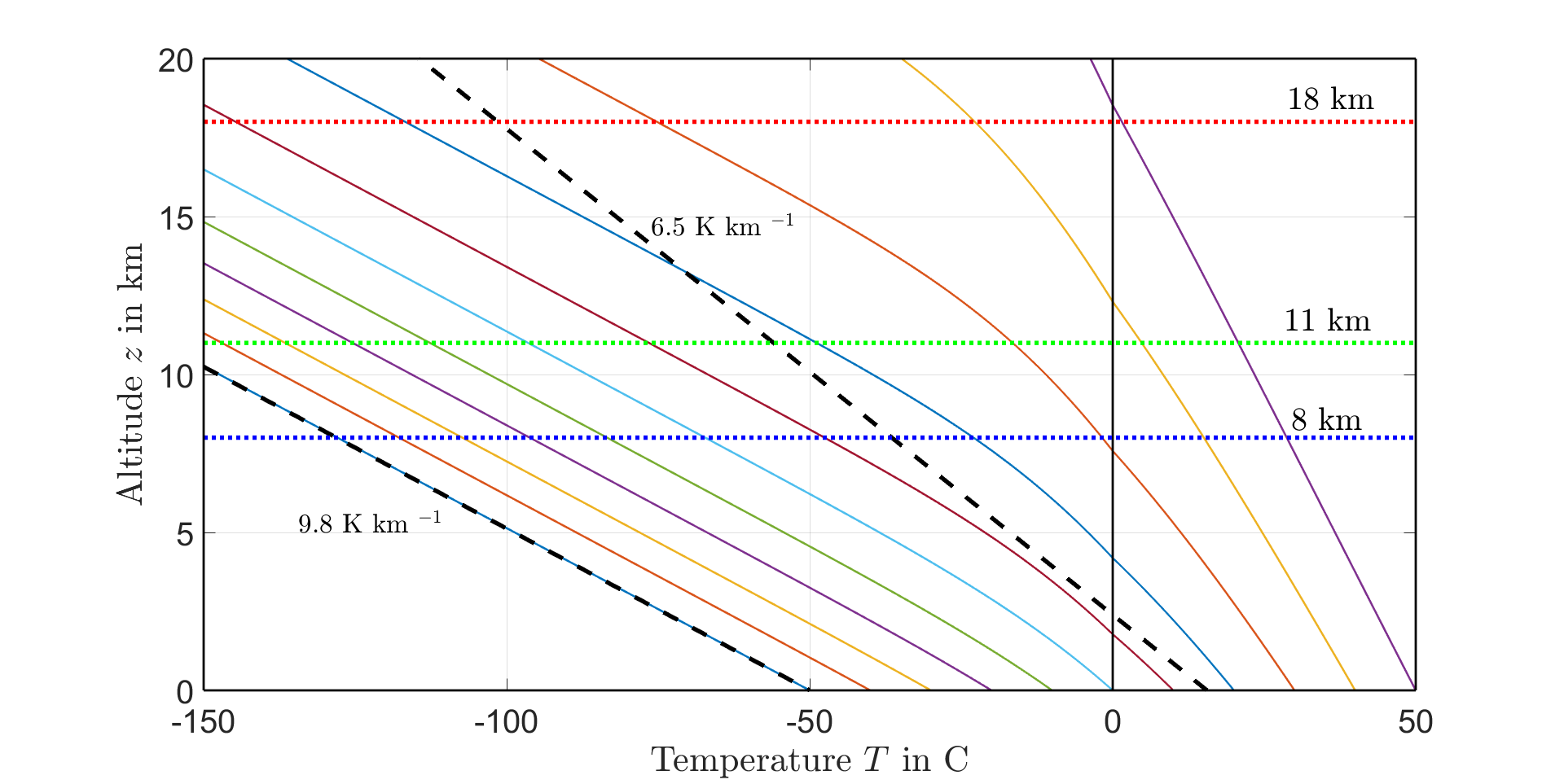} 
\caption{Temperature profiles for lapse rates that are altitude-independent (black dashed lines) and pseudoadiabatic (colored continuous lines).  The dashed black line that starts at a surface temperature of -50 C, has an altitude-independent lapse rate of $\Gamma_W=9.8$ K/km, the dry adiabatic lapse rate of (87).  Little condensible water is available to slow the cooling along the pseudoadiabat that starts from a surface temperature of -50 C, so the dry adiabat and the pseudoadiabat are almost indistinguishable.  The dashed black line that starts from a surface temperature of 15.5 C (288.7 K) has an altitude-independent lapse rate of 6.5 K/km.  Representative tropopause altitudes are 8 km (polar latitudes), 11 km (temperature latitudes) and 18 km (tropical latitudes). 
\label{MALR0}}
\end{figure}

Pseudoadiabatic temperature profiles result if air with 100\% relative humidity is adiabatically decompressed, with liquid or solid phases of water removed as they form.  But no heat is exchanged with the environment during the decompression.  Fig. \ref{MALR0} shows various pseudoadiabatic temperature profiles that can be found using the Clausius Clapeyron equation \cite {Clausius}.  Also shown as black, dashed lines are representative profiles with altitude-independent lapse rates.  The pseudoadiabat with a surface temperature of $\theta_0= -50$ C has very little water vapor and cannot be distinguished from a profile with a fully adiabatic, dry-air lapse rate of 9.8 K/km.

\subsection{Temperature adjustment}
\begin{figure}[t]
\includegraphics[height=80mm,width=0.9\columnwidth]{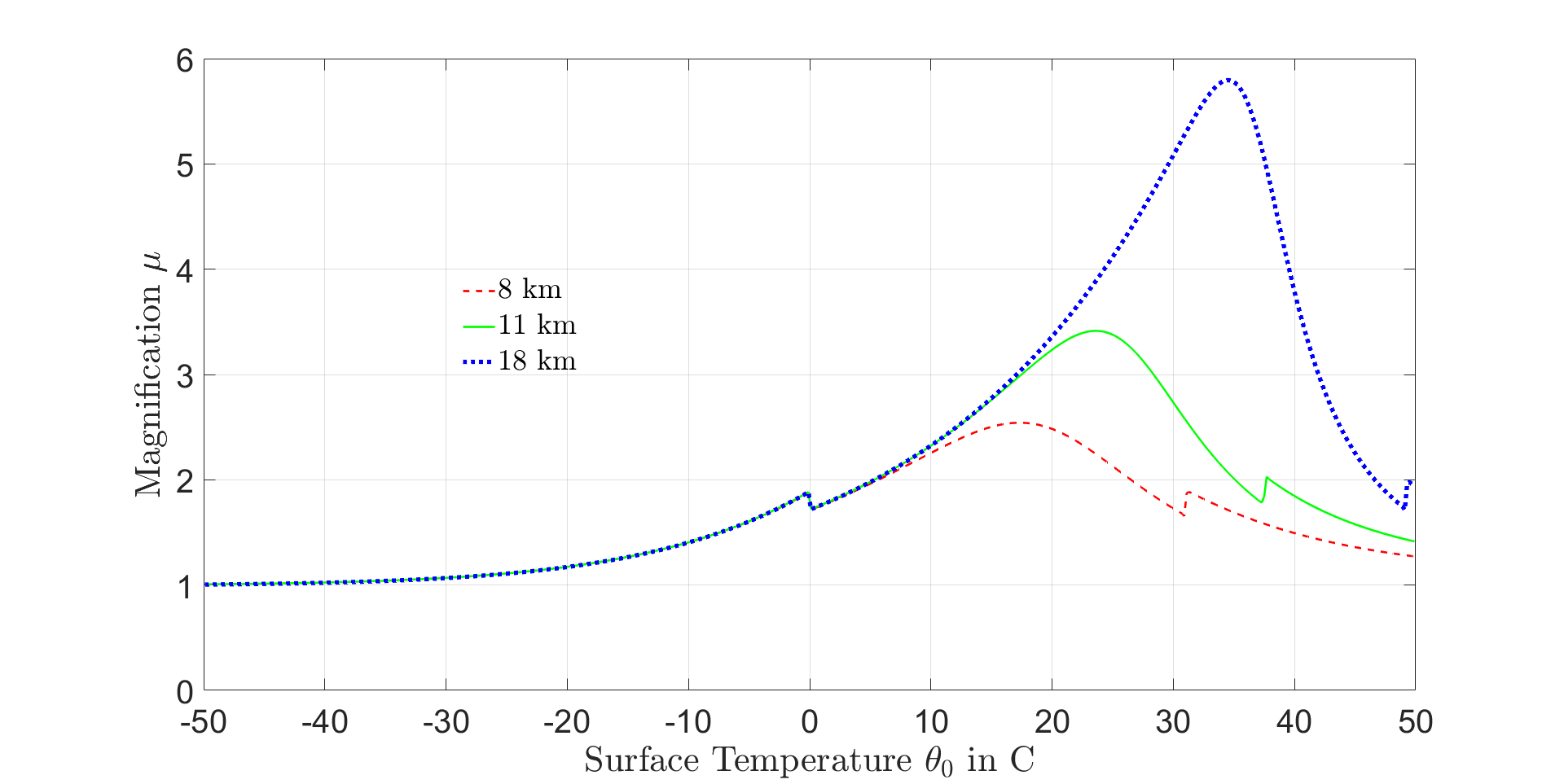} 
\caption{Magnification factor $\mu$ versus Surface Temperature for three tropopause altitudes as indicated.  The left kink occurs at the surface freezing temperature while the right kink occurs when the tropopause temperature $\theta_1=0$ C. 
\label{MAG0}}
\end{figure}

Setting $z=\zeta_1$ in (\ref{rce12}) we see that the breakpoint temperature at the tropopause is
\begin{equation}
\theta_1=T(\zeta_1,\theta_0) = \theta_0-\int_0^{\zeta_1} dz' L(z',\theta_0).
\label{rce14}
\end{equation}
If the surface temperature increases by a small increment $\Delta\theta_0$, the tropopause breakpoint temperature will increase by
$\Delta \theta_1=\mu\Delta\theta_0$.  From (\ref{rce14}) we see that the tropopause temperature magnification factor
\begin{equation}
\mu=\frac{\partial\theta_1}{\partial\theta_0}=1-\frac{\partial }{\partial \theta_0}\int_0^{\zeta_1} dz' L(z',\theta_0).
\label{rce16}
\end{equation}
Some illustrative magnification factors $\mu$ for pseudoadiabatic lapse rates are shown in Fig. \ref{MAG0}.    

From (\ref{rce12}) we see that a surface temperature increase $\Delta\theta_0$, will cause the
tropospheric temperature to increase by $\Delta T $, where
\begin{equation}
\Delta T= \mu\Phi_1\Delta \theta_0=\Phi_1\Delta \theta_1.
\label{rce18}
\end{equation}
The temperature-adjustment basis function $\Phi_1(z)$ is
\begin{equation}
\Phi_1(z)=\frac{1}{\mu}\frac{\partial T}{\partial\theta_0}=\frac{1}{\mu}\left(1-\frac{\partial}{\partial \theta_0}\int_0^z dz' L(z',\theta_0)\right) ,\quad \hbox{for}\quad  z\le \zeta_1.
\label{rce20}
\end{equation}
For altitudes above the tropopause we extend the definition (\ref{rce20}) to
\begin{equation}
\Phi_{1}(z)=\left \{\begin{array}{ll}(\zeta_{2}-z)/(\zeta_{2}-\zeta_{1})\quad&\mbox{if\quad $\zeta_{1}<z<\zeta_{2} $},\\
0\quad &\mbox{if\quad $\zeta_{2}<z $.}\end{array}\right .
\label{rce24}
\end{equation}

\begin{figure}[t]
\includegraphics[height=100mm,width=1.0\columnwidth]{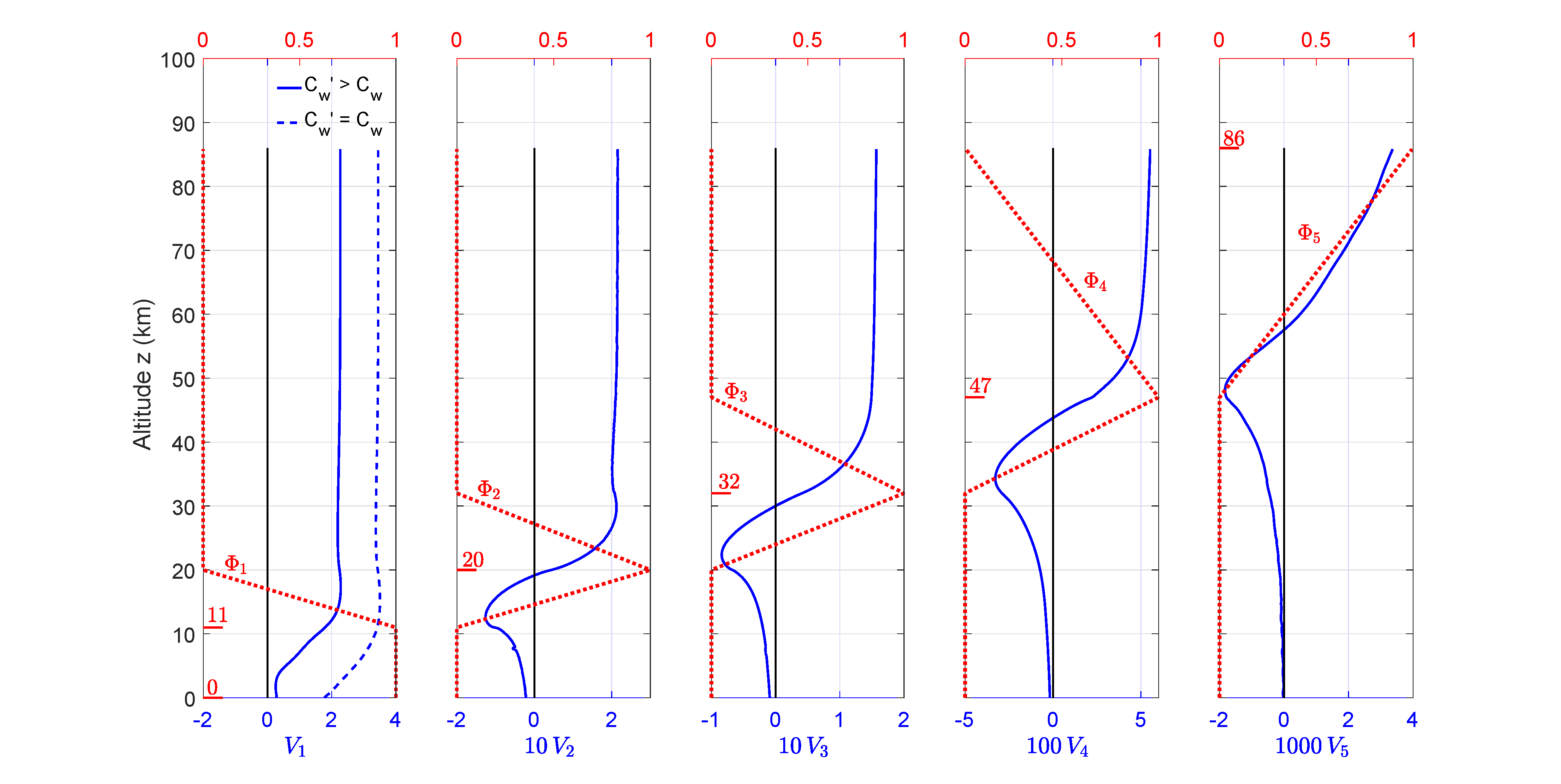} 
\caption{The temperature basis functions $\Phi_{\alpha}$ are shown as dashed red lines with scales at the tops of the panels.
Altitudes at the maxima of the temperature perturbations are given in km next to the short red lines.  Multiples of the flux basis functions $V_{\alpha}$ in units of W m$^{-2}$ K$^{-1}$, and with constant relative humidity water vapor feedback, are shown as solid blue lines, with scales at the bottoms of the panels.  The dashed blue line is for flux perturbation $V_1$ without water vapor feedback (with $C_w'=C_w$). The flux perturbations $V_2,\ldots,V_5$ are the same with or without water vapor feedback.
\label{Vn}}
\end{figure}

Guided by (\ref{rce18}) we will assume that the temperature adjustment $\Delta T$ that best restores radiative-convective equilibrium from the surface to the top of the atmosphere is parameterized by the breakpoint temperature adjustments $\Delta \theta_{\lambda}$, for $\lambda =1,2,3,4,5$ .  We represent the temperature adjustment by the expansion on temperature-adjustment basis functions $\Phi_{\lambda}$.
\begin{equation}
\Delta T =\sum_{\lambda=1}^5\Delta \theta_{\lambda}\Phi_{\lambda}.
\label{rce26}
\end{equation}
We already defined $\Phi_1$ with (\ref{rce20}) and (\ref{rce24}).
For  $\lambda = 2,3,4$ we write the temperature-adjustment basis functions as
\begin{equation}
\Phi_{\lambda}=\left \{\begin{array}{ll}(\zeta_{\lambda+1}-z)/(\zeta_{\lambda+1}-\zeta_{\lambda})\quad&\mbox{if\quad $\zeta_{\lambda}<z<\zeta_{\lambda+1} $},\\
(z-\zeta_{\lambda-1})/(\zeta_{\lambda}-\zeta_{\lambda-1})\quad&\mbox{if\quad $\zeta_{\lambda-1}<z<\zeta_{\lambda} $},\\
0\quad &\mbox{otherwise.}\end{array}\right .
\label{rce38}
\end{equation}
For the highest altitudes we take the temperature-adjustment basis function to be
\begin{equation}
\Phi_{5}=\left \{\begin{array}{ll}(z-\zeta_{4})/(\zeta_{5}-\zeta_{4})\quad&\mbox{if\quad $\zeta_{4}<z $},\\
0\quad &\mbox{otherwise.}\end{array}\right .
\label{rce40}
\end{equation}
Representative temperature-adjustment profiles, $\Phi_{\lambda}$ and flux-adjustment profiles,$V_{\lambda}$  for an altitude-independent lapse rate, $L=6.5$ K km$^{-1}$ are shown in Fig. \ref{Vn}.

\subsection{Flux adjustments}
The breakpoint-temperature adjustments, $\Delta\theta_{\lambda}$, can be used to find the flux that restores radiative-convective equilibrium as

\begin{equation}
Z(C_g', C_w',T') = Z(C_g', C_w,T)+\sum_{\lambda}\Delta\theta_{\lambda}V_{\lambda}.
\label{rce56}
\end{equation}
The flux-adjustment basis vector for $\lambda =1$ is
\begin{equation}
V_{1}=\lim_{\Delta \theta_{1}\to 0}\frac{Z(C_g', C_w', T+\Delta \theta_{1}\Phi_1)-Z(C_g',C_w,T)}{\Delta \theta_{1}}.
\label{rce58}
\end{equation}
Manabe and Wetherald \cite{Manabe1967} pointed out that one plausible way for atmospheric water vapor to respond to a temperature increase would be for the relative humidity to remain constant at all altitudes.  Observational support for this hypothesis is ambiguous.  For example, over 70 years of radiosonde observations indicate relative humidity has decreased slightly in the upper atmosphere \cite{Paltridge}.

We assume that changes in water-vapor concentration have negligible effect at altitudes above the troposophere, so that the flux-adjustment basis vectors for $\lambda =2,3,4,5$
are defined by
\begin{equation}
V_{\lambda}=\lim_{\Delta \theta_{\lambda}\to 0}\frac{Z(C_g', C_w, T+\Delta \theta_{\lambda}\Phi_{\lambda})-Z(C_g',C_w,T)}{\Delta \theta_{\lambda}}.
\label{rce60}
\end{equation}
Then the adjusted forcing of (\ref{rce8}) is

\begin{equation}
\delta Z =\Delta Z+\sum_{\lambda}\Delta\theta_{\lambda}V_{\lambda}.
\label{rce62}
\end{equation}

\subsection{Optimum temperature adjustments}
Between breakpoint altitudes $\zeta_{\lambda}$ and $\zeta_{\lambda+1}$ we have taken 100 evenly spaced altitude segments.
Using (\ref{rce56}) in (\ref{rce8}) for each of the 400 altitude samples $z$ above the tropopause would give 400 linear equations in the five unknowns, $\Delta \theta_1, \Delta\theta_2,\ldots \Delta \theta_5$ . This grossly overdetermines the $\Delta \theta_{\lambda}$. However, we can find values of $\Delta \theta_{\lambda}$ that give the best approximate solution to (\ref{rce8}) by minimizing
\begin{equation}
Q=\sum_{i=1}^{500} W_i (\delta Z_i)^2.
\label{rce64}
\end{equation}
\noindent where $\delta Z_i = \delta Z(z_i)$.  The adjustments $\Delta \theta_{\lambda}$ are not very sensitive to the weights $W_i$, and we used
\begin{equation}
W_i=\left \{\begin{array} {ll} \Delta z_i &\mbox{if } z_i\ge \zeta_1,\\
0 &\mbox{if } z_i<\zeta_1. \end{array}\right .
\label{rce66}
\end{equation}
The altitude interval size is $\Delta z_i = z_{i+1} - z_i$.

The temperature adjustments  $\Delta \theta_{\lambda}$, that minimize (\ref{rce64}) are the simultaneous solutions of the five linear equations $(\lambda = 1,2,3,4,5$)
\begin{equation}
\frac{\partial Q}{\partial \Delta \theta_{\lambda}}=2\sum_i V_{\lambda  i}W_i\, \delta Z_i=0,
\label{rce68}
\end{equation}
where  $V_{\lambda i}=V_\lambda(z_i)$ was defined by (\ref{rce58}) and (\ref{rce60}).  We can write (\ref{rce62}) as the $5\times 5$ matrix equation
\begin{equation}
\sum_{\lambda }A_{\kappa \lambda}\Delta \theta_{\lambda}= \Delta S_{\kappa}.
\label{rce70}
\end{equation}
The adjustment matrix of (\ref{rce70}) is
\begin{equation}
A_{\kappa \lambda}=\sum_i W_i V_{\kappa i}V_{\lambda i}.
\label{rce72}
\end{equation}
The source vector is
\begin{equation}
\Delta S_{\kappa}=-\sum_i \Delta Z_i W_i V_{\kappa i},
\label{rce74}
\end{equation}
where $\Delta Z_i=\Delta Z(z_i)$ was defined by (\ref{rce2}).
We assume that the adjustment matrix $A$ of (\ref{rce72})  has an inverse $A^{-1}$.
Then we can multiply both sides of (\ref{rce70}) by $A^{-1}$ to find
\begin{equation}
\Delta \theta_{\lambda}= \sum_{\kappa} A^{-1}_{\lambda \kappa}\Delta S_{\kappa}.
\label{rce76}
\end{equation}

The temperature adjustment (\ref{rce76})  will slightly affect the breakpoint altitudes.  The pressure will not change for the air segment initially at the altitude $z_i$ since it is due to the weight per unit area of all higher air segments.  The ideal gas law implies that the height increment of the $i$th interval becomes
\begin{equation}
\Delta z'_i=\Delta z_i \frac{T'(z_i)}{T(z_i)}.
\label{rce78}
\end{equation}
The adjusted breakpoint altitudes at the top of the $\lambda$th layer, with $\lambda =1,2,3,4,5$, will be
\begin{equation}
\Delta \zeta_{\lambda}=\sum_{i=1}^{100\lambda}\Delta z'_i.
\label{rce80}
\end{equation}
The surface altitude adjustment obviously is $\Delta \zeta_0 = 0$.
\subsection{Representative Calculations}
\subsubsection{Anderson H$_2$O profile for $L= 6.5$ K km$^{-1}$ }

For the temperature profile and standard greenhouse-gas concentrations of Fig. \ref{GGNT}, doubling the CO$_2$ concentration changes the flux $Z$ as shown in the right panel of Fig. \ref{TempFluxAdjust}.  The adjustments given by (\ref{rce76}) and (\ref{rce80}) are as follows.
\paragraph{Temperature adjustments but no water-vapor adjustment.}
\begin{equation}
\Delta \theta=\left[\begin{array}{r}  1.4\\ 1.4\\ -2.0\\ -7.2\\ -7.9\\ -2.0\end{array}\right]\hbox{ K,}\quad\hbox{and}\quad {\Delta \zeta}=\left[\begin{array}{r}  0\\ 0.06\\ 0.05\\ -0.19\\ -0.65\\ -1.10 \end{array}\right]\hbox{km},\quad\hbox{for $\Delta C_w=0$}.
\label{rr2}
\end{equation}
The breakpoint temperature and altitude adjustments show the lower atmosphere warms and expands slightly after doubling the CO$_2$ concentration while the upper atmosphere cools and contracts.
\paragraph{Both temperature and constant relative humidity water-vapor adjustments.}

\begin{equation}
{\Delta \theta}=\left[\begin{array}{r}  2.3\\ 2.3\\ -2.8\\ -7.0\\ -8.6\\ 3.8\end{array}\right]\hbox{ K},\quad\hbox{and}\quad
{\Delta \zeta}=\left[\begin{array}{r} 0\\ 0.10 \\0.09 \\-0.17\\ -0.64\\-0.98 \end{array}\right]\hbox{km}.
\label{rr4}
\end{equation}
The temperature and altitude adjustments of (\ref{rr4}) are shown in the left panel of Fig. \ref{TempFluxAdjust}.  Doubling CO$_2$ concentrations with water vapor feedback increases the surface temperature warming to $\Delta \theta_0 = 2.3$ K from $\Delta \theta_0 = 1.4$ K, or by a factor of $1.6$.

\subsubsection{Manabe H$_2$O profile for $L= 6.5 \hbox { K km}^{-1}$ }
The sensitivity of the results to the water vapor profile were checked by considering the altitude dependence of relative humidity $\Phi(z)$, used by Manabe and Wetherald \cite{Manabe1967}.  

\begin{equation}
\Phi=\Phi_s \left(\frac{p/p_s - 0.02 }{1-0.02}\right)
\label{rr6}
\end{equation}

\noindent Here, $p$ is the pressure at altitude $z$ while $p_s$ denotes the surface pressure.  Manabe and Wetherald set the surface relative humidity $\Phi_s$ to 77\%.  Equation (\ref{rr6}) fails at high altitudes where it gives negative relative humidity values.  The minimum H$_2$O mixing ratio was therefore set to $3 \times 10^{-6}$ gm per gm of air which corresponds to a concentration of 4.8 ppm. 
Equation (\ref{rr6}) gives a higher surface H$_2$O concentration of 13,396 ppm than 7,750 ppm found using the H$_2$O profile observed by  Anderson \cite{Anderson} shown in Fig. 1.  The H$_2$O concentration obtained using equation (\ref{rr6}) decreases faster with altitude than that observed by Anderson and corresponds to a column density about 20\% higher than that given in Table 1. 

The adjustments given by (\ref{rce76}) and (\ref{rce80}) are as follows.

\paragraph{Temperature adjustments but no water-vapor adjustment.}

\begin{equation}
{\Delta \theta}=\left[\begin{array}{r}  1.4\\ 1.4\\ -1.8\\ -6.7\\ -8.9\\ 0.5\end{array}\right]\hbox{ K},\quad\hbox{and}\quad
{\Delta \zeta}=\left[\begin{array}{r} 0\\ 0.06 \\0.05 \\-0.17\\ -0.64\\-1.33 \end{array}\right]\hbox{km},\quad\hbox{for $\Delta C_w=0$}.
\label{rr8}
\end{equation}

\paragraph{Both temperature and constant relative humidity water-vapor adjustments.}

\begin{equation}
{\Delta \theta}=\left[\begin{array}{r}  2.2\\ 2.2\\ -2.7\\ -6.8\\ -10.0\\ 2.8\end{array}\right]\hbox{ K},\quad\hbox{and}\quad
{\Delta \zeta}=\left[\begin{array}{r} 0\\ 0.10 \\0.09 \\-0.17\\ -0.67\\-1.23 \end{array}\right]\hbox{km}.
\label{mw3}
\end{equation}
These results for both cases without and with water vapor feedback differ very little from (\ref{rr2}) and (\ref{rr4}) that were obtained using the Anderson H$_2$O profile.

\subsubsection{Anderson H$_2$O profile with Pseudoadiabatic Lapse Rate}

The effect of a temperature profile determined using a pseudoadiabatic lapse rate illustrated in Fig. \ref{MALR0} was considered.  For a surface at 288.7 K, the temperature decreases to 211.2 K at 11 km altitude.  This is slightly lower than the tropopause breakpoint temperature $\theta_2 = 217.2$ K given by (\ref{es1}).  The higher altitude temperatures were determined using the same breakpoint temperatures as given by (1) and all of the breakpoint altitudes remained unchanged.  

The adjustments given by (\ref{rce76}) and (\ref{rce80}) are as follows.

\paragraph{Temperature adjustments but no water-vapor adjustment.} 

\begin{equation}
{\Delta \theta}=\left[\begin{array}{r}  1.0\\ 3.0\\ -2.0\\-7.7\\ -8.1\\ 2.1\end{array}\right]\hbox{ K},\quad\hbox{and}\quad
{\Delta \zeta}=\left[\begin{array}{r} 0 \\0.13 \\0.15\\ -0.11\\-0.58 \\-1.05 \end{array}\right]\hbox{km},\quad\hbox{for $\Delta C_w=0$}.
\label{mw2b}
\end{equation}

\paragraph{Both temperature and constant relative humidity water-vapor adjustments.}

\begin{equation}
{\Delta \theta}=\left[\begin{array}{r}  2.2\\ 6.2\\ -4.2\\ -7.3\\ -10.2\\ 8.9\end{array}\right]\hbox{ K},\quad\hbox{and}\quad
{\Delta \zeta}=\left[\begin{array}{r} 0\\ 0.28 \\0.32 \\0.01\\ -0.52\\-0.53 \end{array}\right]\hbox{km}.
\label{mw3b}
\end{equation}
In (\ref{mw2b}) and (\ref{mw3b}), the surface warming was found using the magnification factor $\mu=3.0$ at a midlatitude tropopause altitude of 11 km. 

\subsection{Climate Sensitivity}

A comparison of our result for the climate sensitivity defined as the surface warming due to doubling the CO$_2$ concentration from 400 to 800 ppm, to other work is given in Table 5.  These calculations considered the case of a clear sky one dimensional atmosphere in radiative-convective equilibrium.  All groups get nearly the identical value for the case of fixed absolute humidity for a constant lapse rate of 6.5 K/km in the troposphere.  Additional significant surface warming occurs for the case of fixed relative humidity. 
Our result of 2.2 K is substantially lower than the value obtained by the pioneering work of Manabe and Wetherald \cite{Manabe1967} who obviously did not have access to the current line by line information.  In a later publication \cite{Manabe1975} the authors explained that their 1967 result for the surface warming decreased by about 20\% when they replaced their radiation transfer scheme by that used by Rodgers and Walshaw \cite{Rodgers1966} which they felt was superior.  A 20\% reduction of their climate sensitivity result of 2.9 K gives the value of 2.3 K closer to that of the other groups.  Hunt and Wells \cite{Hunt1979} refined the earlier model used by Manabe and Hunt \cite{Manabe1968} to encompass 18 altitudinal levels up to 37.5 km. Kluft et al \cite{Kluft} calculated radiative fluxes using the Rapid Radiative Transfer Model which is used for global climate models \cite{Mlawer}.  Their temperature adjustments show stratospheric cooling of about 10 K similar to our results but greater surface warming.  All groups obtain similar surface warming for the case of fixed relative humidity using a pseudoadiabatic lapse rate in the troposphere.  Some variation of the results is to be expected since the calculations used different water vapor concentration profiles as well as temperature profiles that differ slightly near the tropopause.

\begin{table}
\begin{center}
\begin{tabular}{|l| c | c | c | c |}
\hline
&&&& \\
\ \ \ \ \ Model Configuration  &Manabe et al 
&Hunt et al &Kluft et al &This Work  \\ [0.5ex]
&\cite{Manabe1967}\cite{Manabe1975} &\cite{Hunt1979} &\cite{Kluft} & \\
&&&& \\
\hline\hline
Fixed absolute humidity, &1.4 (1.4) & &1.3 &1.4 \\
constant lapse rate (6.5 K km$^{-1}$) &&&& \\
\hline
Fixed relative humidity, &2.9 (2.2) &2.2 &2.7 &2.3 \\
constant lapse rate (6.5 K km$^{-1}$) &&&& \\
\hline
Fixed relative humidity, &2.0 &1.8 &2.1 &2.2 \\
pseudoadiabatic lapse rate &&&& \\
\hline
\end{tabular}
\end{center}

\caption{Climate sensitivity in Kelvins for different model configurations.  The bracketed numbers next to the results found by Manabe and Wetherald \cite{Manabe1967} are the results of our calculation using their relative humidity profile given by (\ref{rr6}).  Our result for the case of fixed relative humidity with a pseudoadiabatic lapse rate in the troposphere was found using the temperature profile shown in Fig. (\ref{MALR0}) which is further discussed in the text.}
\label{Table5}
\end{table}

\section{Comparison of Model Intensities to Satellite Observations\label{ld}}

\begin{figure}[t]
\includegraphics[height=100mm,width=1.0\columnwidth]{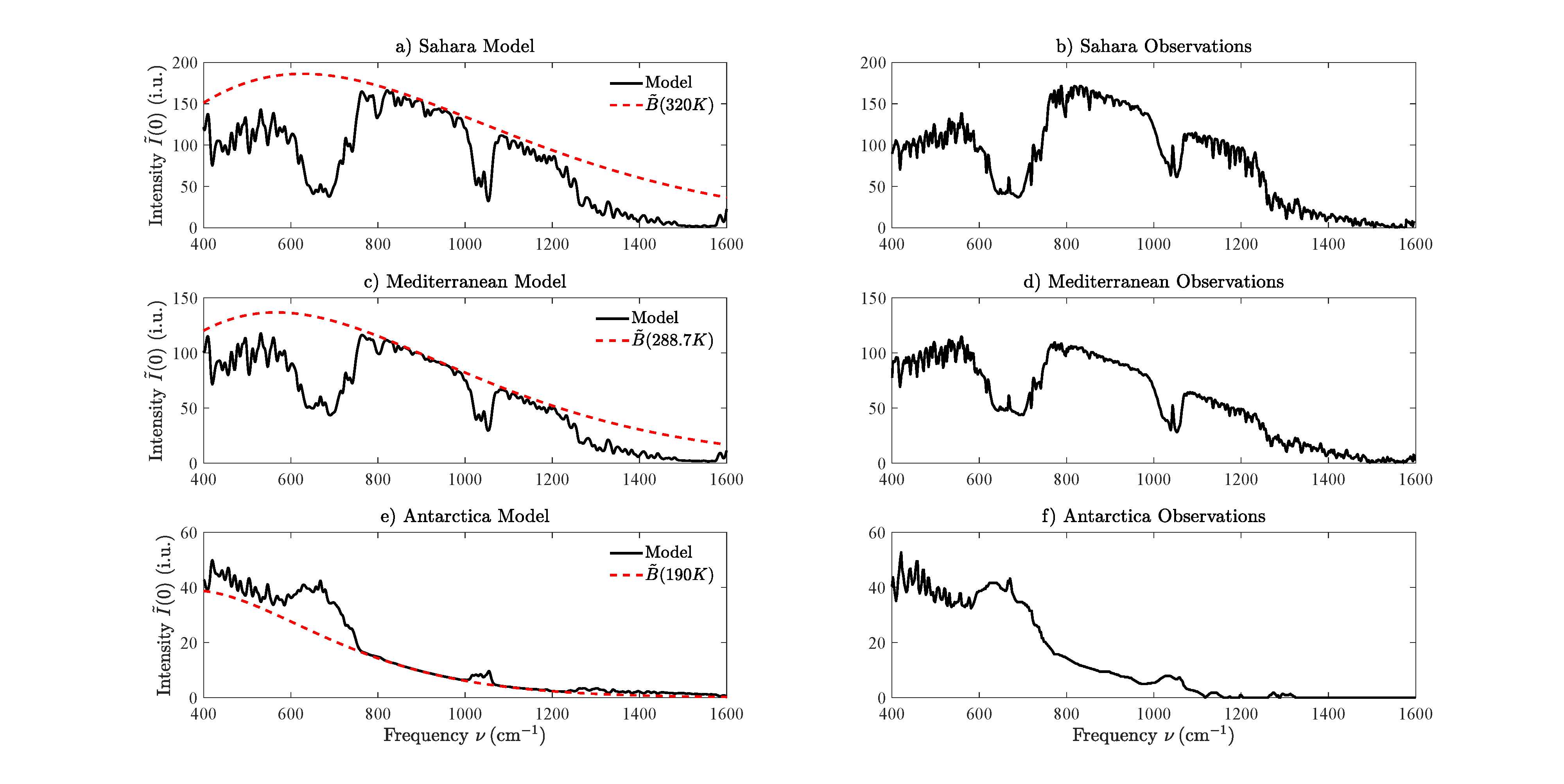}
\caption{Vertical intensities $\tilde I(0)$ at the top of the atmosphere observed with a Michaelson interferometer in a satellite \cite{Nimbus}, and modeled with (\ref{vn48}): over the Sahara desert, the Mediterranean and  Antarctica. The intensity unit is 1 i.u. = 1 mW m$^{-2}$ cm sr$^{-1}$. Radiative forcing is negative over wintertime Antarctica since the relatively warm greenhouse gases in the troposphere, mostly CO$_2$, O$_3$ and H$_2$O, radiate more to space than the cold ice surface, at a temperature of $T=190$ K, could radiate through a transparent atmosphere.
\label{Nimbus}}
\end{figure}

An important test is to compare calculations to observations.  Fig \ref{Nimbus} shows vertical spectral intensities, $\tilde I(0)$, measured with a Michaelson interferometer from a satellite over the Sahara Desert, the Mediterranean Sea and Antarctica \cite{Nimbus}. The figure also shows values of the vertical intensity, $\tilde I$, calculated with (\ref{vn48}).

For the Mediterranean, we used the five segment temperature profile of Fig. \ref{GGNT}.
For the Sahara and Antarctica analogous profiles with different parameters were used.  The altitude breakpoints for the Sahara were at $\zeta=[0,18,20,32,47,86]$ km.  The high tropopause at 18 km is characteristic of near equatorial latitudes.  For Antarctica, the altitude breakpoints were $\zeta=[0, 2.5, 8, 25, 47, 86]$. The low tropopause at 8 km is characteristic of the nighttime poles, as is the strong, wintertime temperature inversion, peaking at 2.5 km.  The lapse rates between the breakpoints were  $L = [6.5,0,-1,-3.8, 2.145]$ K km$^{-1}$ for the Sahara and $L = [-12.5, 2.33,  0,-1.5, 2.145]$ K km$^{-1}$ for Antarctica.  The surface temperature in the Sahara was taken to be $T_0=320$ K (very hot) and the surface temperature in Antarctica  was taken to be $T_0 = 190$ K (very cold).  The surface pressure in the Sahara and the Mediterranean was taken to be $p_0 = 1013$ hPa and the surface pressure in Antarctic was taken to be $p_0= 677$ hPa, low because of the high elevation of the ice surface, about $2.7$ km above mean sea level.

For convenience, we modeled the dependence of the water vapor concentrations $C^{\{i\}}$ on the height $z$ above the surface as

\begin{equation}	
C^{\{i\}}=C_0^{\{i\}} e^{-z/z_w},
\label{ld2}
\end{equation}
with a latitude independent scale height $z_w=5$ km and with surface concentrations $C_0^{\{i\}}=31,000$ ppm for the Sahara, $C_0^{\{i\}}=12,000$ ppm for the Mediterranean,
and $C_0^{\{i\}}=2,000$ ppm  Antarctica.  For the year 1970 when the satellite measurements were made, we used surface concentrations,
in ppm, for CO$_2$, N$_2$O and CH$_4$ of 326, 0.294 and 1.4, with the same relative altitude profile as those in Fig \ref{GGNT}.  The altitude profile of Fig. \ref{GGNT} for O$_3$ was used for the Sahara, the Mediterranean and Antarctica.

As can be seen from Fig. \ref{Nimbus} the modeled spectral intensities can hardly be distinguished from the observed values.  We conclude that our modeled spectral fluxes would also be close to observed fluxes, if a reliable way to measure spectral fluxes were invented.

\section{Conclusions}

This work examined the transmission of infrared radiation through a cloud-free atmosphere from the Earth's surface to outer space.  A line by line calculation used over $1/3$ million lines of the five most important naturally occurring greenhouse gases, H$_2$O, CO$_2$, O$_3$, N$_2$O and CH$_4$.  This included considerably more weaker rovibrational line strengths, for H$_2$O as small as $10^{-27}$ cm, than other studies.  The calculation of forcings took into account the observed altitudinal concentrations of the various gases as well as several temperature profiles.

The upward spectral flux, $\tilde Z$, ``breaks out" at an emission height $z_e$, given by (\ref{am0}). Emission heights can be near the top of the atmosphere for frequencies in the middle of strong absorption lines.  For frequencies with little absorption, the emission heights can be close to, or at the surface as shown in Fig. \ref{tZ500-1016}.

The most striking fact about radiation transfer in Earth's atmosphere is summarized by Figs. \ref{CO2} and \ref{CH4}.  Doubling the current concentrations of the greenhouse gases CO$_2$, N$_2$O and CH$_4$  increases the forcings by a few percent for cloud-free parts of the atmosphere.
Table \ref{int2} shows the forcings at both the top of the atmosphere and at the tropopause are comparable to those found by other groups.  

Radiative forcing depends strongly on latitude, as shown in Figs. \ref{ZzCM} and \ref{ZzCA}. Near the wintertime poles, with very little water vapor in the atmosphere, CO$_2$ dominates the radiative forcing. The radiation to space from H$_2$O, CO$_2$ and O$_3$ in the relatively warm upper atmosphere can exceed the radiation from the cold surface of the ice sheet and the TOA forcing can be negative.

Fig. \ref{DFDC} as well as Tables \ref{int} and \ref{dPr} show that at current concentrations, the forcings from all greenhouse gases are saturated.  The saturations of the abundant greenhouse gases H$_2$O and CO$_2$ are so extreme that the per-molecule forcing is attenuated by four orders of magnitude with respect to the optically thin values.  Saturation also suppresses the forcing power per molecule for the less abundant greenhouse gases, O$_3$, N$_2$O and CH$_4$, from their optically thin values, but far less than for H$_2$O and CO$_2$.

Table \ref{int} and Fig. \ref{Pbar} show the overlap of absorption bands of greenhouse gases causes their forcings to be only roughly additive.  One greenhouse gas interferes with, and diminishes, the forcings of all others.  But the self-interference of a greenhouse gas with itself, or saturation, is a much larger effect than interference between different gases.  Table \ref{dPr} shows that for optically thin conditions, the forcing power per molecule is about the same for all greenhouse gases, a few times $10^{-22}$ W per molecule.

Doubling the CO$_2$ concentration will cause a temperature decrease of the upper atmosphere of about 10 K as shown in Fig. \ref{TempFluxAdjust} to restore hypothetical radiative-convective equilibrium.  For the case of fixed absolute humidity, the surface warms by 1.4 K which agrees very well with other work as shown in Table \ref{Table5}.  The surface warming increases significantly for the case of water feedback assuming fixed relative humidity.  Our result of 2.3 K is within 0.1 K of values obtained by two other groups as well as a separate calculation where we used the Manabe water vapor profile given by (\ref{rr6}).  For the case of fixed relative humidity and a pseudoadiabatic lapse rate in the troposphere, we obtain a climate sensitivity of 2.2 K.  The corresponding climate sensitivities determined by other groups differ by about 10\% which can be expected using slightly differing temperature and water vapor profiles.  The issue of water feedback would undoubtedly be greatly clarified if additional observations of water vapor concentration as a function of altitude were available.  

Fig. \ref{Nimbus} shows that the integral transform (\ref{vn48}) used to calculate TOA intensities $\tilde I$ with HITRAN line intensities and with no CO$_2$ nor H$_2$O continuum absorption gives results in very close agreement with spectral intensities observed from satellites over climate zones as different as the Sahara Desert, the Mediterranean Sea and Antarctica.  One can therefore have confidence in the calculations of spectral fluxes.  The negligible effect of the H$_2$O continuum on the top of the atmosphere radiative flux has also been found by Zhong and Haigh \cite{Zhong}.  It would be interesting to examine comparable data for the tropics where atmospheric moisture is highest to determine the effect of a H$_2$O continuum.  One would need to be careful that any ``observed continuum" not be confused with a layer of cloud like haze which can be prevalent at high humidities.
In conclusion, the combination of one dimensional radiative-convective models and observations such as TOA intensities are invaluable for furthering our understanding of how increasing greenhouse gas concentrations will affect the Earth's climate.

\section*{Acknowledgements}
We are grateful for constructive suggestions by many colleagues. Special thanks are due to Tom Sheahen for initial encouragement and to G. Iouli who helped access the HITRAN data base.  The Canadian Natural Science and Engineering Research  Council provided financial support of one of us.


\begin{thebibliography}{99}
\bibitem{IPCC} G. Myhre et al, {\it Anthropogenic and Natural Radiative Forcing}, {\it Climate Change 2013:  The Physical Science Basis.  Contribution of Working Group I to the Fifth Assessment Report of the Intergovernmental Panel on Climate Change}. Cambridge University Press, Cambridge, United Kingdom (2013).

\bibitem{MaunaLoa} Global Greenhouse Gas Reference Network, NOAA Earth System Research Laboratory/Global Monitoring Division, www.esrl.noaa.gov/gmc/ccgg, (2019).

\bibitem{Schwartz1}  S. E. Schwartz, {\it Resource Letter GECC-1:  The Greenhouse Effect and Climate Change:  Earth's Natural Greenhouse Effect}, Am. J. Phys. {\bf 86}, (8), 565-576, (2018).

\bibitem{Schwartz2} S. E. Schwartz, {\it Resource Letter GECC-2:  The Greenhouse Effect and Climate Change:  The Intensified Greenhouse Effect},  Am. J. Phys. {\bf 86}, (9), 645-656, (2018).

\bibitem{Rothman92} L. S. Rothman, R. L. Hawkins, R. B. Watson and R. R. Gemache, {\it Energy Levels, Intensities and Linewidths of Atmospheric Carbon Dioxide Bands}, JQSRT, {\bf 48}, 537 (1992).

\bibitem{HITRAN} I. E. Gordon, L. S. Rothman et al., {\it The HITRAN2016 Molecular Spectrospic Database}, JQSRT {\bf 203}, 3-69 (2017).

\bibitem{Chandrasekhar} S. Chandrasekhar, {\it Radiative Transfer}, Dover, New York (1960).

\bibitem{Manabe1961} S. Manabe and F. M\"oller, {\it On the Radiation Equilibrium and Heat Balance of the Atmosphere}, Monthly Weather Review {\bf 89}, No. 12, 503 (1961).

\bibitem{Goody} R. M. Goody and Y. L. Yung, {\it Atmospheric Radiation: Theoretical Basis} Oxford University Press, Oxford (1995).

\bibitem{Edwards1992} D. P. Edwards, {\it A General Line-by-Line Atmospheric Transmittance and Radiance Model}, NCAR Technical Note, NCAR/TN-367+STR (1992).

\bibitem{Clough1992} S. A. Clough, M. J. Iacono and J. L. Moncet, {\it Line-by-Line Calculations of Atmospheric Fluxes and Cooling Rates: Applications to Water Vapor}, J. Geophys. Res. {\bf 97}, 15761 (1992).

\bibitem{Myhre1998} G. Myhre, E. J. Highwood, K. P. Shine and F. Stordal, {\it New Estimates of Radiative Forcing due to Well Mixed Greenhouse Gases}, Geophys. Res. Lett. {\bf 25}, 2715 (1998).

\bibitem{Collins2006} W. D. Collins {\it et al.} {\it Radiative Forcing by Well Mixed Greenhouse Gases:  Estimates from Climate Models in the Intergovernmental Panel on Climate Change (IPCC) Fourth Assessment Report (AR4)}, J. Geophys. Res. {\bf 111}, D14317 (2006).

\bibitem{Harde2013} H. Harde, {\it Radiative and Heat Transfer in the Atmosphere: A Comprehensive Approach on a Molecular Basis}, Int. J. Atm. Sci. 503727, (2013).

\bibitem{Schreier2014} F. Schreier, S. G. Garcia, P. Hedelt, M. Hess, J. Mendrok, M. Vasquez and J. Xu, {\it GARLIC -- A General Purpose Atmospheric Radiation Transfer Line-by-Line Infrared-Microwave Code: Implementation and Evaluation}, JQSRT {\bf 137}, 29 (2014).

\bibitem{Etminan2016} M. Etminan, G. Myhre, E. J. Highwood and K. P. Shine, {\it Radiative Forcing of Carbon Dioxide, Methane and Nitrous Oxide: A Significant Revision of the Methane Radiative Forcing}, Geophys. Res. Lett. {\bf 43}, 12614 (2016).

\bibitem{Temp} {\it The U.S. Standard Atmosphere}, NASA Report TM-X-74335 (1976).

\bibitem{Anderson} G. P. Anderson, S. A. Clough, F. X. Kneizys, J. H. Chetwynd and E. P. Shettle, {\it AFGL Atmospheric Constituent Profiles (0-120 km)}, AFGL-TR-86-0110  Air Force Geophysics Laboratory, Hanscom Air Force Base, Massachusetts, (1986).

\bibitem{Einstein} A. Einstein, {\it Zur Quantentheorie der Strahlung}, Physik. Zeitschrift {\bf 18}, 121 (1917).
	
\bibitem{Dirac1930} P. A. M. Dirac, {\it The Principles of Quantum Mechanics}, Oxford University Press, New York (1930).

\bibitem{Voigt12} W. Voigt, {\it On the Intensity Distribution Within Lines of a Gaseous Spectrum}, Sitzungsber. Math.-Phys. Kl. Bayer. Akad. Wis., M\"unchen, p. 603 (1912).
	
\bibitem{Schwarzkopf85} M. D. Schwarzkopf and S. B. Fels, {\it Improvements of the Algorithm for Computing CO$_2$ Transmissivities and Cooling Rates}, J. Geophys. Res. {\bf 90},10541-10550, (1985).

\bibitem{Lacis91} A. A. Lacis and V. Oinas, {\it A Description of the Correlated k Distribution Method for Modeling Nongray Gaseous Absorption, Thermal Emission, and Multiple Scattering in Vertically Inhomogeneous Atmospheres}, J. Geophys. Res. {\bf 96}, 9027-9063, (1991).
	
\bibitem{Edwards1991} D. P. Edwards and L. L. Strow, {\it Spectral Lineshape Considerations for limb Temperature Sounders}, J. Geophys. Res. {\bf 96}, 20859 (1991).

\bibitem{Schwarzschild1906} K. Schwarzschild, {\it \"Uber Diffusion und Absorption in der Sonnenatmosph\"are}, Nachr. K. Gesell. Wiss. Math.-Phys. Klasse {\bf 195}, 41 (1906).

\bibitem{Buglia} J. J. Buglia, {\it Introduction to the Theory of Atmospheric Radiative Transfer}, NASA Publication 1156 (1986).

\bibitem{Rodgers1966} C. D. Rodgers and C. D. Walshaw, {\it The computation of infra-red cooling rate in planetary atmospheres}, Quart. J. R. Met. Soc. {\bf 92}, 67 (1966).

\bibitem{Wilber} A. C. Wilber, D. P. Kratz and S. K. Gupta, {\it Surface Emissivity Maps for Use in Satellite Retrievals of Longwave Radiation}, NASA/TP-1999-209362 (1999).

\bibitem{Yoshikawa} K. K. Yoshikawa, {\it An Iterative Solution of an Integral Equation for Radiative Transfer using Variational Technique}, NASA Technical Report TN-D-7292, A-4774 (1973).

\bibitem{Abramowitz} M. Abramowitz and I. Stegun, {\it Handbook of Mathematical Functions}, Dover Publications, New York (1965).

\bibitem{Smithuesen} H. Smith\"usen, J. Notholt, G. K\"onig-Langlo, P. Lemke and T. Jung, {\it How Increasing CO$_2$ leads to an Increased Negative Greenhouse Effect in Antarctica}, Geophys. Res. Lett. {\bf 42}, 10422 (2015).

\bibitem{Zhong2013} W. Zhong and J. D. Haigh, {\it The Greenhouse Effect and Carbon Dioxide}, Weather {\bf 68}, 100 (2013).

\bibitem{Arrhenius1908} S. Arrhenius, {\it  Worlds in the Making; the Evolution of the Universe}, translated  by Dr. H. Borns, Harper, New York, London (1908).

\bibitem{Wilson12} D. J. Wilson and J. Gea-Banacloche, {\it Simple Model to Estimate the Contribution of Atmospheric CO$_2$ to the Earth's Greenhouse Effect}, Am. J. Phys. {\bf 80} 306 (2012).

\bibitem{Manabe1967} S. Manabe and R. T. Wetherald, {\it Thermal Equilibrium of the Atmosphere with a Given Distribution of Relative Humidity}, J. Atmos. Sci. {\bf 24}, 241 (1967).

\bibitem{Radiosondes} J. Nash, {\it Measurement of the Upper-Air Pressure, Temperature and Humidity}, Instruments and Observing Methods, Report No. 121. World Meteor. Org. (2015).

\bibitem{Clausius} R. Clausius, {\it Mechanisache W\"armetheorie}, Friedrich Vieweg und Sohn, Braunschweig, Vol. 1, Section VI-11. (1876).

\bibitem{Paltridge} G. Paltridge, A. Arking and M. Pook, {\it Trends in middle- and upper-level tropospheric humidity from NCEP reanalysis data}, Theor. Appl. Climatol. {\bf 98}, 351 (2009).

\bibitem{Manabe1975} S. Manabe and R. T. Wetherald, {\it The Effects of Doubling the CO$_2$ Concentration on the Climate of a General Circulation Model}, J. Atmos. Sci. {\bf 32}, 3 (1975).

\bibitem{Hunt1979} B. G. Hunt and N. C. Wells, {\it An Assessment of the Possible Future Climatic Impact of Carbon Dioxide Increases Based on a Coupled One-Dimensional Atmospheric-Oceanic Model}, J. Geophys. Res. {\bf 84}, 787 (1979).

\bibitem{Manabe1968} S. Manabe and B. G. Hunt, {\it Experiments with a Stratospheric General Circulation Model, Radiative and Dynamic Aspects}, Mon. Weather Rev. {\bf 96}, 477 (1968).

\bibitem{Kluft} L. Kluft, S. Dacie, H. Schmidt and B. Stevens, {\it Re-Examining the First Climate Models:  Climate Sensitivity of a Modern Radiative-Conective Equilibrium Model}, J. Climate {\bf 32}, 8111 (2019).

\bibitem{Mlawer} E. J. Mlawer, S. J. Taubman, P. D. Brown, M. J. Iacono and S. A. Clough, {\it Radiative Transfer for Inhomogeneous Atmospheres:  RRTM, a Validated Correlated-k Model for the Longwave}, J. Geophys. Res. {\bf 102}, 16,663 (1997).

\bibitem{Nimbus} R. A. Hanel and B. J. Conrath, {\it Thermal Emission Spectra of the Earth and Atmosphere from the Nimbus 4 Michaelson Interferometer Experiment}, Nature {\bf 228}, 143 (1970).

\bibitem{Zhong} W. Zhong and J. D. Haigh, {\it Improved Broadband Emissivity Parameterization for Water Vapor Cooling Rate Calculations}, J. Atmos. Sci. {\bf 52} 124 (1995).

\end{thebibliography}
\end{document}